\documentclass[a4paper,10pt]{article}

\usepackage{amssymb,amsbsy}
\usepackage{amsmath,amscd}
\usepackage{amssymb}
\usepackage{amsmath,tabu}
\usepackage{amsfonts}
\usepackage{color}
\usepackage{mathrsfs, graphicx}

\newcommand\cH{{\mathcal H}}
\newcommand\cR{{\mathcal R}}
\newcommand\cL{{\mathcal L}}
\newcommand\bbbC{{\mathbb C}}
\newcommand\bbbR{{\mathbb R}}

\newcommand\bx{{\bf x}}

\newtheorem{The}{Theorem}
\newtheorem{Lem}{Lemma}
\newtheorem{Cor}{Corollary}
\newtheorem{Rem}{Remark}
\newtheorem{Def}{Definition}
\newtheorem{Ex}{Example}

\begin{document}
\bibliographystyle{unsrt} 


\author{V. M. Buchstaber and A. V. Mikhailov}

\title{The space
of symmetric squares of hyperelliptic curves 
and integrable Hamiltonian polynomial systems on $\bbbR^4$}

\maketitle

\begin{abstract}
We construct Lie algebras of vector fields on universal bundles 
$\mathcal{E}^2_{N,0}$ of symmetric squares
of hyperelliptic curves of genus $g=1,2,\dots$, where 
$g=\left[\frac{N-1}{2}\right], \ N=3,4,\ldots$.
For each of these Lie algebras, the Lie subalgebra of vertical
fields has commuting generators,
while the generators of the Lie subalgebra of projectable fields
determines
the canonical representation of the Lie subalgebra with
generators $L_{2q}$, $q=-1, 0, 1, 2, \dots$, of the Witt algebra.
We give explicitly a bi-rational equivalence of the space $\mathcal{E}^2_{N,0}$ 
and $\bbbC^{N+1}$ (in the case $N=5$ it is a well known result of Dubrovin 
and Novikov) and construct a polynomial Lie algebra on $\bbbC^{N+1}$, 
which contains two 
commuting generators. These commuting generators results in two compatible 
polynomial dynamical systems on $\bbbR^4$, which possess two common polynomial 
first integrals. Moreover, these systems are Hamiltonian and thus 
Liouville integrable. Using Abel-Jacobi two point map the solutions of these 
systems can be given in terms of 
functions defined on universal  covering of the universal bundle of 
the Jacobians of the curves. These functions are not Abelian if $g\ne 2$. 
Finally we give explicit solutions of the constructed Hamiltonian systems on 
$\bbbR^4$ in the cases $N=3,4,5$.

\end{abstract}

\section*{Introduction}

This is an extended version of our paper \cite{BM-2017}.
In this article we obtain a description of the Lie algebras
$\mathcal{G}(\mathcal{E}^2_{N,0})$
of vector fields
on the spaces of the universal bundles
$\mathcal{E}^2_{N,0}$ (see Definition~2)
of the \emph{symmetric squares} \emph{of the hyperelliptic curves}
$$
V_{\bx}=\bigg\{ (X,Y)\in \mathbb{C}^2:Y^2=\prod_{i=1}^N (X-x_i)\bigg\},\qquad \bx=(x_1,\dots,x_N).
$$
The Lie algebra
$\mathcal{G}(\mathcal{E}^2_{N,0})$ contains the Lie subalgebra of fields lifted from the
base $\mbox{Sym}^N(\mathbb{C})$, i.e. 
\emph{horizontal} and
\emph{projectable} fields (see~\cite{R1} and~\cite[p.~337]{R2}) over
the polynomial Lie algebra $\mathcal{G}(\mbox{Sym}^N(\mathbb{C}))$
(see \cite{Arn-90}, \cite{BL-02-3}) of derivations of the ring of symmetric 
polynomials
in $x_1,\dots,x_N$.

The Lie algebra $\mathcal{G}(\mbox{Sym}^N(\mathbb{C}))$ naturally arises and 
plays an important role
in various areas of mathematics and mathematical physics,
including the isospectral deformation method \cite{perelomov_book} and the 
classical method of 
separation of variables \cite{tsyganov_book}.
In fundamental works (see, e.g.,
\cite{Arn-90} and
\cite{Arn}) as coordinates on $\mbox{Sym}^N(\mathbb{C})$
the elementary symmetric functions $e_1,\dots,e_N$ were chosen.
In \cite{Arn-90}, in terms of the action of the permutation group $S_N$
on $\mathbb{C}^N$,
the operation of convolution of invariants was introduced
and basis vector fields on $\mbox{Sym}^N(\mathbb{C})$ were defined, which
are independent at any point of the variety of regular orbits.
At each point of the variety of irregular orbits
these fields generate the tangent space
to the stratum of the discriminant hypersurface containing the given point.
Zakalyukin's well-known construction yields basis vector fields
$V_i=\sum V_{i,j}(\mathbf{e})\frac{\partial}{\partial e_j}$, 
$\mathbf{e}=(e_1,\dots,e_N)$,
with \emph{symmetric} matrix $V_{i,j}$, which are tangent to the discriminant.

In this article we show that the use of the Newton polynomials
$p_1,\dots,p_N$ makes it possible to substantially simplify
the formulas and, most importantly,
employ the remarkable
infinite-dimensional Lie subalgebra $W_{-1}$ of the Witt algebra $W$ in 
computations.
The generators of the Lie algebra $W_{-1}$ are
$L_{-2},L_0,L_2,\dots$,
and the commutation relations have the form 
$[L_{2q_1},L_{2q_2}]=2(q_2-q_1)L_{2(q_1+q_2)}$.

For any $N$, the Lie algebra $W_{-1}$ has a faithful canonical representation
in the Lie algebra $\mathcal{G}(\mbox{Sym}^N(\mathbb{C}))$ which maps the 
generator $L_{2q}$ to
the Newton field $\mathcal{L}_{2q}^0=2\sum x_i^{q+1}\partial_{x_i}$.
The image of this representation belongs to the Lie algebra $W_{-1}(N)$, which 
has the structure of a free left module over the polynomial
ring $\mathbb{C}(\mbox{Sym}^N(\mathbb{C}))$.
We obtain an explicit expression for the vector fields $V_i$
with symmetric matrix $V_{i,j}(\mathbf{e})$
in terms of the Newton fields $\mathcal{L}_{2q}^0$ (see Corollary~\ref{cor-7}).
An important role in our calculations is played by a
grading of variables and operators.
In this connection, we introduce
variables $y_{2m}$ and $\mathcal{N}_{2k}$ by setting $e_m=y_{2m}$ and 
$p_k=\mathcal{N}_{2k}$ and bear in mind that
$\deg x_i=2$.

Note that the Jacobi identity in the
$\mathbb{C}[\mathcal{N}_2,\dots,\mathcal{N}_{2N}]$-polynomial Lie algebra 
$W_{-1}(N)$
implies the nontrivial differential relation
$$
\sum_{m=1}^N m\bigg(\mathcal{N}_{2(k+m)}\frac{\partial 
\mathcal{N}_{2(q+n)}}{\partial \mathcal{N}_{2m}} 
-\mathcal{N}_{2(q+m)}\frac{\partial \mathcal{N}_{2(k+n)}}{\partial 
\mathcal{N}_{2m}} \bigg) =
(q-k)\mathcal{N}_{2(k+q+n)}
$$
for the Newton polynomials.
We show that there exists a \emph{unique} representation of the Lie algebra 
$W_{-1}$
in the Lie algebra of \emph{horizontal} vector fields $L_{-2},L_0,L_2,\dots$
of the Lie algebroid
$\mathcal{G}(\mathcal{E}^2_{N,0})$
(see Theorem~\ref{t-6}). The proof of the uniqueness of this representation uses 
the fact
that, relative to the Lie bracket $[\cdot,\cdot]$,
the Lie algebra $W_{-1}$ is generated by  $L_{-2}$ and $L_4$
satisfying the relation $[L_2,[L_2,[L_2,L_4]]]=12[L_4,[L_2,L_4]]$.

On the Lie algebroid $\mathcal{G}(\mathcal{E}^2_{N,0})$
there exist two commuting \emph{vertical} vector fields.
For each point $\bx$, the restrictions of these fields to $\mbox{Sym}^2(V_\bx)$
are the images of obviously commuting fields on $V_\bx\times V_\bx$.
In our upcoming publication we shall give an explicit description
of such commuting fields on $\mbox{Sym}^k(V_\bx)$ for any $k\ge 2$.
In Section~4 of this article we give an explicit description
of these fields in the case of interest to us, namely, for $k=2$.

Similar operators
on $\mbox{Sym}^m(\mathbb{C}^2)$ were constructed in \cite{Enr-Rub-03}
on the basis of a construction of the spectral curve and the Poisson structure.
A proof of the commutativity of operators
in \cite{Enr-Rub-03} uses a method that differs from ours.

The key results of our work are a formula for the generating
series $L(t)$ (see Theorem~\ref{t-7})
of the horizontal vector fields 
$\mathcal{L}_{-2}^0,\dots,\mathcal{L}_{2k}^0,\dots$
that determine a representation of the Lie algebra $W_{-1}$
and a commutation formula for the vertical and horizontal vector fields
(see Theorem \ref{t-8}).

In Section~6 we construct an $N$-dimensional algebraic variety $W(N)$
in the $(N+2)$--dimensional algebraic variety $\mathcal{E}^2_{N,0}$
and homeomorphisms $f_N \colon \mathbb{C}^{N+2}\setminus\{u_4=0\}\to
\mathcal{E}^2_{N,0}\setminus W(N)$,
where $\{u_4=0\}$ is a hyperplane in $\mathbb{C}^{N+2}$ with graded coordinates 
$u_2,u_4,v_{N-2},v_N;y_2,\dots,y_{2(N-2)}$. In the partial case $N=5$ when we 
consider a universal space of symmetric squares of hyperelliptic curves of 
genus 2 we a obtain a new proof of a well known result obtained by Dubrovin and 
Novikov (see section 4 in the review paper \cite{DMN}).  
One of the main results of the present work is an explicit construction,
which uses the homeomorphisms $f_N$,
of the polynomial Lie algebras on $\mathbb{C}^{N+2}$ that are determined
by the Lie algebroids $\mathcal{G}(\mathcal{E}^2_{N,0})$, $N=3,4,\dots$
(see Theorem \ref{t-9}).

In Section \ref{s10} we give explicit description of integrable Hamiltonian 
polynomial systems in $\bbbR^4$, associated with $\mathcal{E}^2_{N,0}$. In the 
cases $N=3,4,5,7$ represent the systems, their Hamiltonians and solutions.

\section{The Space of Symmetric Squares of Hyperelliptic Curves}

Consider a family of plane curves
\begin{equation}\label{c-1}
V_{N,0}=\{ (X,Y;\bx)\in \mathbb{C}^2\times \mathbb{C}^N:\pi(X,Y;\bx)=0 \},
\end{equation}
where
\begin{equation}\label{pi-2}
\pi(X,Y;\bx)=Y^2-\prod_{k=1}^N(X-x_k).
\end{equation}
The vector $\xi_N(\bx)=\mathbf{e}$ is the \emph{parameter set} for a curve
in family (\ref{c-1}).
By $V_{N}$ we denote the subfamily of curves satisfying $e_1=p_1=0$.

In this paper we use the following grading of variables:
$\deg x_k=2$, $k=1,\dots,N$, $\deg X=2$, and $\deg Y=N$.
With respect to this grading the polynomial $\pi(X,Y;\bx)$
is homogeneous of degree~$2N$.

The \emph{discriminant variety} of family (\ref{c-1})
is the algebraic variety
$$
\mbox{Disc}(V_{N,0})=\{\xi_N(\bx)\in \mbox{Sym}^N(\mathbb{C}):\Delta_N=0 
\},\quad\text{where }\,\Delta_N=\prod\limits_{i<j}(x_i-x_j)^2.
$$
The discriminant variety $\mbox{Disc}(V_{N})$ of the family of curves $V_{N}$ is 
defined similarly.
The variety $\mbox{Disc}(V_{N,0})\subset \mathbb{C}^N$ is the image under the 
projection
$\mathbb{C}^N\to\mbox{Sym}^N(\mathbb{C})=\mathbb{C}^N$ of the union of
the so-called mirrors, that is, the hyperplanes $\{x_i=x_j,\,i\ne j\}$, and 
$\mbox{Disc}(V_{N})\subset \mathbb{C}^{N-1}$
is the image of the intersection of the space
$\mathbb{C}^{N-1}=\{{\bf e}\in\mathbb{C}^n:e_1=0\}$ with the union of mirrors.

For $N=3$, the variety $\mbox{Disc}(V_{3})\subset \mathbb{C}^{2}$ in the 
coordinates
$(e_2,e_3)$
is determined by the equation $\Delta_3 = 27 e_3^2 - 4 e_2^3 =0$, i.e.,
is the well-known swallowtail
in $\mathbb{C}^2$. In the book \cite{Arn-90} it was proposed
to refer to the varieties $\mbox{Disc}(V_{N})$
as generalized swallowtails in $\mathbb{C}^{N-1}$.

Let $\mathcal{B}_{N,0}$ and $\mathcal{B}_{N}$ denote
the open varieties $\mathbb{C}^N\setminus
\mbox{Disc}(V_{N,0})$ and $\mathbb{C}^{N-1}\setminus \mbox{Disc}(V_{N})$, 
respectively.
The curves of the families $V_{N,0}$ and $V_{N}$
with parameters in the spaces $\mathcal{B}_{N,0}$ and $\mathcal{B}_{N}$
are said to be nonsingular for obvious reasons.
They have genus
$\left[ \frac{N-1}{2} \right]$.
For example, in the cases $N=3$ and $4$, these are elliptic curves.

We set $\widehat{\mathcal{E}}_{N,0}=\{ (X,Y;\bx)\in \mathbb{C}^2\times
\mathbb{C}^N: \pi(X,Y;\bx)=0, \,\xi(\bx)\in \mathcal{B}_{N,0} \}$.
The group $S_N$ acts freely on $\widehat{\mathcal{E}}_{N,0}$ by permutations of 
the coordinates
$x_1,\dots,x_N$, and therefore a regular $N!$-sheeted covering
$\widehat{\mathcal{E}}_{N,0}\to\mathcal{E}_{N,0}= 
\widehat{\mathcal{E}}_{N,0}/S_N$ is defined.

\begin{Def}\label{def-1}
The \textit{universal bundle of nonsingular hyperelliptic curves of family}
(\ref{c-1})
is the bundle $\mathcal{E}_{N,0}\to \mathcal{B}_{N,0}\colon 
(X,Y;\xi(\bx))\mapsto \xi(\bx)$.
\end{Def}

The space $\mathcal{E}_{N,0}$ is the \textit{universal space} of nonsingular
hyperelliptic curves of genus $\left[ \frac{N-1}{2} \right]$.
The fiber over a point of the base $\mathcal{B}_{N,0}$ is the curve with 
parameters determined
by this point. The universal bundle $\mathcal{E}_{N}\to\mathcal{B}_{N}$ of 
nonsingular hyperelliptic curves
is defined similarly.

We set
$$
\widehat{\mathcal{E}}_{N,0}^2=\{ (X_1,Y_1;X_2,Y_2;\bx)\in 
\mathbb{C}^2\times\mathbb{C}^2\times \mathbb{C}^N:
\pi(X_k,Y_k;\bx)=0, \, k=1,2;\, X_1-X_2\neq 0, \,\xi(\bx)\in \mathcal{B}_{N,0} 
\}.
$$
The group $G=S_2\times S_N$ acts freely on $\widehat{\mathcal{E}}_{N,0}^2$, so 
that
the generator of $S_2$ determines the permutation $(X_1,Y_1)\leftrightarrow 
(X_2,Y_2)$,
and the elements of the group $S_N$
determine permutations of the coordinates of the vector $\bx$. Therefore,
a regular covering
$\mathcal{E}_{N,0}^2=\widehat{\mathcal{E}}_{N,0}^2/G \to 
\widehat{\mathcal{E}}_{N,0}^2$ is defined.

\begin{Def}\label{def-2}
The \textit{universal bundle of symmetric squares of nonsingular hyperelliptic
curves of family} (\ref{c-1})
is the bundle
$$
\mathcal{E}_{N,0}^2\to\mathcal{B}_{N,0} \colon ([X_1,Y_1;X_2,Y_2];[\bx])\mapsto 
[\bx],
$$
where $[X_1,Y_1;X_2,Y_2]=\xi_2(X_1,Y_1;X_2,Y_2)$, $[\bx]=\xi(\bx)$, and $\xi_2 
\colon
\mathbb{C}^2\times\mathbb{C}^2 \to\mbox{Sym}^2(\mathbb{C}^2)$ is the canonical 
projection
onto the symmetric square of $\mathbb{C}^2$.
\end{Def}

The space $\mathcal{E}_{N,0}^2$ is called the \emph{universal space} of 
symmetric squares of
nonsingular hyperelliptic curves of genus
$\left[ \frac{N-1}{2} \right]$.
The fiber over a point of the base $\mathcal{B}_{N,0}$
is the variety $(\mbox{Sym}^2V)\setminus(X_1-X_2=0)$
with parameters determined by this point.

The universal bundle $\mathcal{E}_{N}^2\to\mathcal{B}_{N}$ is defined similarly.

\section{The Lie Algebra of Newton Vector Fields}

In the paper \cite{BL-02-3} the theory of polynomial Lie algebras was 
constructed.
Important examples of such infinite-dimensional Lie algebras are the Lie 
algebras of vector fields
on $\mathbb{C}^N$ and $\mathbb{C}^{N-1}$ tangent to the varieties 
$\mbox{Disc}(V_{N,0})$
and $\mbox{Disc}(V_{N})$ and,
therefore, the Lie algebras of vector fields on $\mathcal{B}_{N,0}$ and 
$\mathcal{B}_{N}$.

In this section we give an explicit description of the Lie algebras
$\mathcal{G}_{P_0}(N)$ and $\mathcal{G}_P(N)$
of vector fields in coordinates $(p_1,\dots,p_N)$ and $(p_2,\dots,p_N)$ 
determined by the Newton polynomials.
We have $\deg x_i=2$, $i=1,\dots,N$.
To control grading, we
introduce the notation
$$
\mathcal{N}_{2k}=p_k(\bx)=\sum_{i=1}^N x_i^k, \qquad k=0,1,\dots\,.
$$
For graded generators of the polynomial ring
$\mathcal{P}(\mbox{Sym}^N(\mathbb{C}))$
we take the polynomials $\mathcal{N}_{2},\dots,\mathcal{N}_{2N}$. Then
$\mathcal{P}(\mbox{Sym}^N(\mathbb{C}))\simeq 
\mathbb{C}[\mathcal{N}_{2},\dots,\mathcal{N}_{2N}]$.

\begin{Def}\label{def-3}
The gradient homogeneous polynomial vector fields
\begin{equation}\label{f-2}
\mathcal{L}_{2q}^0=2\sum_{i=1}^N x_i^{q+1}\partial_{x_i},\qquad q=-1,0,1,\dots,
\end{equation}
on $\mathbb{C}^N$ of degree $2q$ are called the \textit{Newton derivations} of 
the ring $\mathbb{C}[x_1,\dots,x_N]$.
\end{Def}

\begin{Lem}\label{L-1}
The operators $\mathcal{L}_{2q}^0$, $q=-1,0,1,\dots$
are derivations of the ring 
$\mathbb{C}[\mathcal{N}_{2},\dots,\mathcal{N}_{2N}]$,
and they are uniquely determined by the formula
\begin{equation}\label{f-5}
\mathcal{L}_{2q}^0\mathcal{N}_{2k}=2k\mathcal{N}_{2(k+q)}, \qquad k=1,2,\dots\,.
\end{equation}
\end{Lem}

\begin{Cor}\label{cor-1}
The operators
$\mathcal{L}_{2q}^0$ act on the ring 
$\mathbb{C}[\mathcal{N}_{2},\dots,\mathcal{N}_{2N}]$
as \vglue-13pt
\begin{equation}\label{f-6}
\mathcal{L}_{2q}^0=\sum_{k=1}^N 2k\mathcal{N}_{2(q+k)}\,\frac{\partial}{\partial 
\mathcal{N}_{2k}}\,.
\end{equation}
\end{Cor}

Lemma \ref{L-1} and Corollary \ref{cor-1} are verified directly.

Let us write the equation $\prod_{i=1}^N (x-x_i)=0$
in the form $x^N=\sum_{j=1}^N (-1)^{j+1} y_{2j}x^{N-j}$.

\begin{Lem}\label{L-2}
For any $k\ge 0$,
\begin{equation}\label{f-8}
x^{N+k}=\sum_{j=1}^N (-1)^{j+1} y_{2k,2j}x^{N-j},
\end{equation}
where the $\{y_{2k,2j},\,k=1,\dots\}$, $j=1,\dots,N$, are the sets of symmetric 
functions
in $x_1,\dots,x_N$ with generating series
\begin{equation}\label{f-9}
\mathcal{Y}_{2j}=\sum_{k=0}^\infty 
y_{2k,2j}t^k=\frac{1}{E(t)}\sum_{s=0}^{N-j}(-1)^s y_{2(j+s)}t^s, \qquad 
E(t)=\prod_{i=1}^N (1-x_it).
\end{equation}
\end{Lem}

{\bf Proof}
According to (\ref{f-8}), for any $k\ge 0$,
we have
\begin{align*}
\sum_{j=1}^N (-1)^{j+1} y_{2(k+1),2j}x^{N-j}&=x^{N+k+1}=x\cdot x^{N+k}\\[-10pt]
&=y_{2k,2}\bigg[ \sum_{j=1}^N (-1)^{j+1} y_{2j}x^{N-j}\bigg]\! +\sum_{j=2}^N 
(-1)^{j+1} y_{2k,2j}x^{N-j+1}.
\end{align*}
Hence $y_{2(k+1),2N}= y_{2N} y_{2k,2}$ and $y_{2(k+1),2j}=y_{2j}y_{2k,2}- 
y_{2k,2(j+1)},\ j<N$.
We obtain the following system of equations for the generating series:
\begin{align*}
\mathcal{Y}_{2N} &= y_{2N}(1+t\mathcal{Y}_2), \\
\mathcal{Y}_{2j} &= y_{2j}(1+t\mathcal{Y}_2)-t\mathcal{Y}_{2(j+1)},\qquad 
j=1,\dots,N-1.
\end{align*}
Solving this system, we obtain (\ref{f-9}).\\
\hfill $\Box $

\begin{Cor}\label{cor-2}
For any $k\ge 0$,
\begin{align}
\label{f-10}
\mathcal{N}_{2(N+k)}&=\sum_{j=1}^N (-1)^{j+1} y_{2k,2j}\mathcal{N}_{2(N-j)},\\
\label{f-11}
\mathcal{L}_{2(N+k-1)}^0&=\sum_{j=1}^N (-1)^{j+1} 
y_{2k,2j}\mathcal{L}_{2(N-j-1)}^0.
\end{align}
\end{Cor}

Let us introduce the generating series
$\mathcal{L}^0(t)=\sum_{q=-1}^\infty \mathcal{L}_{2q}^0t^{q+1}$.

\begin{Cor}\label{cor-3}
The following relation holds:
\begin{equation}\label{r-2}
\mathcal{L}^0(t)=\frac{1}{E(t)}\sum_{m=1}^N E(t;m)\mathcal{L}_{2(m-2)}^0 
t^{m-1},
\end{equation}
where
\begin{equation}\label{r-3}
E(t;m)=\sum_{k=0}^{N\!-m}(-1)^k y_{2k}\,t^k \quad\text{and}\quad E(t)=E(t;0).
\end{equation}
\end{Cor}

{\bf Poof} 
Let us write the series $\mathcal{L}^0(t)$ in the form
$\mathcal{L}^0(t)=\sum_{q=-1}^{N\!-2}\mathcal{L}_{2q}^0t^{q+1} + 
t^N\sum_{k=0}^\infty \mathcal{L}_{2(N+k-1)}^0 t^k$.
According to (\ref{f-11}), we obtain
\begin{align*}
\mathcal{L}^0(t)&=\sum_{q=-1}^{N\!-2}\mathcal{L}_{2q}^0t^{q+1} + t^N 
\sum_{j=1}^N (-1)^{j+1} \bigg(\sum_{k=0}^\infty 
y_{2k,2j}\,t^k\bigg)\mathcal{L}_{2(N-j-1)}^0\\
&= \sum_{j=1}^N t^{N\!-j}(1+(-1)^{j+1}t^j\mathcal{Y}_{2j}) 
\mathcal{L}_{2(N-j-1)}^0.
\end{align*}
It remains to use (\ref{f-9}).\\
\hfill $\Box $

\begin{Lem}\label{L-3}
The following relation holds:
\begin{equation}\label{f-12}
[\mathcal{L}_{2q_1}^0,\mathcal{L}_{2q_2}^0]=2(q_2-q_1)\mathcal{L}_{2(q_1+q_2)}
^0.
\end{equation}
\end{Lem}

{\bf Poof} 
The required relation follows directly from (\ref{f-5}).\\
\hfill $\Box $

\begin{Cor}\label{cor-4}
For all $k,q\in \mathbb{N}$ and $n=1,\dots$,
the polynomials $\mathcal{N}_{2k}$, $k=0,1,\dots$,
are related by
\begin{equation}\label{f-13}
\sum_{m=1}^N m\bigg(\mathcal{N}_{2(k+m)}\frac{\partial 
\mathcal{N}_{2(q+n)}}{\partial \mathcal{N}_{2m}} -
\mathcal{N}_{2(q+m)}\frac{\partial \mathcal{N}_{2(k+n)}}{\partial 
\mathcal{N}_{2m}} \bigg) =
(q-k)\mathcal{N}_{2(k+q+n)}.
\end{equation}
\end{Cor}

{\bf Poof} 
The required relations follow directly from
(\ref{f-6}) and (\ref{f-12}).\\
\hfill $\Box $

\begin{Ex}
For $N=3$, the polynomials $\mathcal{N}_2$, $\mathcal{N}_4$, and $\mathcal{N}_6$
are algebraically independent and $\mathcal{N}_0=3$.
For $k=0$, $q=1$, and $n=3$, relation (\ref{f-13})
gives the Euler differential equation
$$
2\mathcal{N}_2\,\frac{\partial \mathcal{N}_8}{\partial \mathcal{N}_2} + 
4\mathcal{N}_4\,
\frac{\partial \mathcal{N}_8}{\partial \mathcal{N}_4} + 
6\mathcal{N}_6\,\frac{\partial
\mathcal{N}_8}{\partial \mathcal{N}_6} =8 \mathcal{N}_8.
$$
\end{Ex}

\begin{Lem}\label{L-4}
For all $q=-1,0,1,\dots$,
\begin{equation}\label{f-14}
\mathcal{L}_{2q}^0\Delta(N)=4\gamma_{2q}(N)\Delta(N),
\end{equation}
where
\begin{equation}\label{f-15}
\Delta(N)=\prod\limits_{i<j}(x_i-x_j)^2\quad\text{and}\quad\gamma_{2q}(N) = 
\sum_{i<j}\frac{x_i^{q+1}-x_j^{q+1}}{x_i-x_j}\,.
\end{equation}
For any $k\ge 0$,
$$
\gamma_{2(N+k)}(N) = \sum_{s=1}^{N-1} (-1)^{s+1} y_{2(k+1),2s}\gamma_{2(N-s-1)}.
$$
\end{Lem}

{\bf Poof} 
We have
$$
\mathcal{L}_{2q}^0\Delta(N)=4\Delta(N)\sum_{k=1}^N 
x_k^{q+1}\partial_k\ln\prod_{i<j}(x_i-x_j)=
4\Delta(N)\bigg(\sum_{i<j}\frac{x_i^{q+1}-x_j^{q+1}}{x_i-x_j} \bigg).
$$
The formula for $\gamma_{2(N+k)}(N)$ follows from (\ref{f-8}).\\
\hfill $\Box $

\begin{Ex}
$\gamma_{-2}(N)=0$, $\gamma_{0}(N)= 2N(N-1)$, and 
$\gamma_{1}(N)=4(N-1)\mathcal{N}_2$.
\end{Ex}

\begin{Cor}\label{cor-5}
The vector fields $\mathcal{L}_{2q}^0$, $q=-1,0,1,\dots$, on $\mathbb{C}^N$
determine vector fields on
$\mbox{Sym}^N(\mathbb{C})$ tangent to the algebraic variety
$\mbox{Disc}(V_{N,0})\subset \mbox{Sym}^N(\mathbb{C})$.
\end{Cor}

\begin{The}\label{t-1}
The Lie algebra $\mathcal{G}_{P_0}(N)$ of vector fields on the variety 
$\mathcal{B}_{N,0}$
in the coordinates $\mathcal{N}_2,\dots,\mathcal{N}_{2N}$ has the structure
of a free $N$-dimensional module
over the ring $\mathbb{C}[\mathcal{N}_2,\dots,\mathcal{N}_{2N}]$
with generators $\mathcal{L}_{2q}^0$, $q=-1,0,1,\dots,N-2$.
The set of generators extends to an infinite set $\{\mathcal{L}_{2q}^0\}$,
where the elements $\mathcal{L}_{2q}^0$ for $q>N-2$ are given by (\ref{f-11}).
The operators $\mathcal{L}_{2q}^0$ act on $\mathcal{N}_{2k}$ by (\ref{f-5}).

The structure of the Lie algebra $\mathcal{G}_{P_0}(N)$
is determined by (\ref{f-12}) and (\ref{f-5}),
where the $\mathcal{N}_{2k}$ for $k>N$ are the polynomials
$\mathcal{N}_{2k}(\mathcal{N}_2,\dots,\mathcal{N}_{2N})$
defined recursively by (\ref{f-10}).
\end{The}

We set $\mathcal{L}_A^0(t)=tE(t)\mathcal{L}^0(t)$.
According to Corollary~\ref{cor-3}, operators
$\mathcal{L}_{A,2(k-2)}^0$
for which
$$
\mathcal{L}_A^0(t)=\sum_{m=1}^N \mathcal{L}_{A,2(m-2)}^0t^m = \sum_{m=1}^N 
E(t;m)\mathcal{L}_{2(m-2)}^0 t^{m}
$$
are defined.

\begin{Lem}\label{L-5}
The following relation holds: $\mathcal{L}_A^0(t)=\sum_{k=1}^N t\prod_{j\neq 
k}(1-x_jt)\frac{\partial}{\partial x_k}$.
\end{Lem}

{\bf Poof}  
We have
$$
\mathcal{L}_A^0(t)=tE(t)\sum_{q=-1}^\infty 
\mathcal{L}_{2q}^0t^{q+1}=E(t)\sum_{k=1}^N 
\frac{t}{1-x_kt}\,\frac{\partial}{\partial x_k} = \sum_{k=1}^N t \prod_{j\neq 
k} 
(1-x_jt)\frac{\partial}{\partial x_k}\,.
$$
\hfill $\Box $

\begin{Ex}
$\mathcal{L}_{A,-2}^0=\mathcal{L}_{-2}^0$ and 
$\mathcal{L}_{A,2(N-2)}^0=y_{2N}\mathcal{L}_{-4}^0$.
\end{Ex}

\begin{Lem}\label{L-6}
The generating polynomial
$$
\mathcal{L}_A^0(t)E(s)=\sum_{m=1}^N(-1)^m (\mathcal{L}_A(t)y_{2m})s^m
$$
in $s$ is symmetric with respect to the permutation $t\leftrightarrow s$.
\end{Lem}

{\bf Poof}  
We have
$$
\mathcal{L}_A^0(t)E(s)=tE(t)\mathcal{L}^0(t)E(s)=-E(t)E(s)\sum_{i=1}^N 
\frac{t}{1-x_it}\,\frac{s}{1-x_is}\,.
$$
\hfill $\Box $

\begin{Cor}\label{cor-6}
The action of the operators $\mathcal{L}_{A,2(k-2)}^0$, $k=1,\dots,N$,
on the elementary symmetric polynomials $e_m=y_{2m}$
is given by a symmetric matrix $T_{k,m}^0=T_{k,m}^0(y_2,\dots,y_{2N})$, that is,
$$
\mathcal{L}_{A,2(k-2)}^0y_{2m} = \mathcal{L}_{A,2(m-2)}^0y_{2k}.
$$
\end{Cor}

{\bf Poof} 
We have
\begin{align*}
\mathcal{L}_A^0(t)E(s)&= \bigg(\sum_{k=1}^N (-1)^k \mathcal{L}_{A,2(k-2)}^0t^k 
\bigg) \bigg(\sum_{m=1}^N(-1)^m y_{2m}s^m \bigg)\\
&= \sum_{k=1}^N \sum_{m=1}^N (-1)^{k+m} (\mathcal{L}_{A,2(k-2)}^0y_{2m})t^ks^m.
\end{align*}
It remains to use Lemma~\ref{L-6}.\\
\hfill $\Box $

The Lie algebra $\mathcal{G}_{P}(N)$ of vector fields on $\mathcal{B}_{N}$
in the coordinates $\mathcal{N}_4,\dots,\mathcal{N}_{2N}$ is a Lie subalgebra
of $\mathcal{G}_{P_0}(N)$.
It consists of the fields that leave
the ideal 
$J_2=\langle\mathcal{N}_2\rangle\subset\mathbb{C}[\mathcal{N}_2,\dots,\mathcal{N
}_{2N}]$
invariant.

We have $\mathcal{L}_{-2}^0\mathcal{N}_2=2N$ and 
$\mathcal{L}_{2q}^0\mathcal{N}_2=2\mathcal{N}_{2(q+1)}$.
We set $\mathcal{L}_{0}=\mathcal{L}_{0}^0$ and $\mathcal{L}_{2q} = 
\mathcal{L}_{2q}^0-\frac{1}{N}\mathcal{N}_{2(q+1)}\mathcal{L}_{-2}^0$.
By construction $\mathcal{L}_{0}\mathcal{N}_2 = 2\mathcal{N}_2$
and $\mathcal{L}_{2q}\mathcal{N}_2=0$ for $q\neq 0$.

\begin{The}\label{t-2}
The Lie algebra $\mathcal{G}_{P}(N)$ of vector fields on $\mathcal{B}_{N}$
in the coordinates $\mathcal{N}_4,\dots,\mathcal{N}_{2N}$ has the structure
of a free $(N-1)$-dimensional module
over the ring $\mathbb{C}[\mathcal{N}_4,\dots,\mathcal{N}_{2N}]$ with generators
$\mathcal{L}_{2q}$, $q=0,1,\dots,N-2$.
The set of generators extends to an infinite set $\{\mathcal{L}_{2q}\}$,
and the elements $\mathcal{L}_{2q}$ for $q=N+k-1$, $k\ge 0$, are given by
(see (\ref{f-11}))
\begin{equation}\label{L2q-1}
\mathcal{L}_{2(N+k-1)}^0=\sum_{j=1}^{N}(-1)^{j+\!1}\widehat{y}_{2k,2j}\mathcal{L
}_{2(N-j-1)}^0,
\end{equation}
where $\widehat y_{2k,2j}\equiv y_{2k,2j}\!\!\mod\!\langle\mathcal{N}_2\rangle$
and the $y_{2k,2j}$ are the polynomials determined by the generating 
series~(\ref{f-9}).

The structure of the Lie algebra on $\mathcal{G}_{P}(N)$
is introduced directly by the condition
that this is a Lie subalgebra of the Lie algebra $\mathcal{G}_{P_0}(N)$.
\end{The}

Let $\mathcal{L}_A(t)=tE(t)\mathcal{L}(t)$, where
$$
\mathcal{L}(t)=\sum_{q=-1}^\infty \mathcal{L}_{2q}t^{q+1}=\sum_{q=-1}^\infty 
\bigg(\mathcal{L}_{2q}-\frac{1}{N}\mathcal{N}_{2(q+1)}\mathcal{L}_{-2}^0 
\bigg)t^{q+1}.
$$

\begin{Lem}\label{LL-7}
The following relation holds:
\begin{equation}\label{2-19}
\mathcal{L}_A(t)=\mathcal{L}_A^0(t) + \frac{1}{N}\bigg(\mathcal{L}_{-2}^0E(t) 
\bigg)\mathcal{L}_{-2}^0 =
\bigg[ \mathcal{L}_{0} -\bigg(1-\frac{4}{N}\bigg)y_2\mathcal{L}_{-2} \bigg]t^2 
+ 
\cdots\,.
\end{equation}
\end{Lem}

Lemmas \ref{L-6} and \ref{LL-7} give the following result.

\begin{Lem}\label{LL-8}
The generating polynomial
$\mathcal{L}_A(t)E(s)=\sum_{m=1}^N(-1)^m (\mathcal{L}_A(t)y_{2m})s^m$
is symmetric with respect to the permutation $t\leftrightarrow s$.
\end{Lem}

Let us introduce operators $\mathcal{L}_{A,2(k-2)}^0$ such that
$\mathcal{L}_A(t) = \sum_{k=2}^N\mathcal{L}_{A,2(k-2)}t$.

\begin{Cor}\label{cor-7}
The operators $\mathcal{L}_{A,2(k-2)}$, $k=2,\dots,N$, leave
the ideal $\langle\mathcal{N}_2\rangle$ invariant.
Their action on the elementary symmetric polynomials $e_m=y_{2m}$ is determined
up to the ideal $\langle\mathcal{N}_2\rangle$
by a symmetric matrix $T_{k,m}=T_{k,m}(y_4,\dots,y_{2N})$.
\end{Cor}

{\bf Poof} 
The proof of this assertion is similar to that
of Corollary~\ref{cor-6}.\\
\hfill $\Box $

The Lie algebra $\mathcal{G}_{P}(N)$ of vector fields on $\mathcal{B}_{N}$
in the coordinates $y_4,\dots,y_{2N}$
has the structure of a free ($N-1$)-dimensional module over the ring
$\mathbb{C}[y_4,\dots,y_{2N}]$
with generators $\mathcal{L}_{A,2(k-2)}$, $k=2,\dots,N$.
The action of these generators on $y_{2m}$ is given by a symmetric matrix
$(T_{k,m})=(T_{k,m}(N))$.

\begin{Rem}
For each $N$, we give an explicit construction
of the fields $\mathcal{L}_{A,2(k-2)}$
and the symmetric matrix $(T_{k,m}(N))$.
The notation $\mathcal{L}_A$ is suggested by
Arnold's monograph \cite{Arn-90} (see also
\cite{Arn}).
These fields will be used in Section~9 to construct the Lie algebroids
$\mathcal{G}(\mathcal{E}^2_{N,0})$
and explicitly describe an isomorphism
between the Lie algebroid $\mathcal{G}(\mathcal{E}^2_{5,0})$
and the algebroid constructed in \cite{B-16} from
the universal bundle of Jacobians of genus $2$ curves.
\end{Rem}

\section{Representations of the Witt Algebra $\boldsymbol{W_{-1}}$ in Lie 
Algebras\\
with the Structure of a Free $\boldsymbol{N}$-Dimensional Module\\ over the 
Polynomial
Ring}

Let us introduce the following notion.

\begin{Def}
We define
an \textit{$N$-polynomial Lie algebra} $W_{-1}(N)$ as the graded Lie algebra 
with

$\bullet$ the structure of a \textit{free left} module
over the graded ring $A(N)=\mathbb{C}[v_2,\dots,v_{2N}]$, $\deg v_{2k}=2k$;

$\bullet$ an infinite set of generators
$L_{2q}^0$, $q=-1,0,1,\dots,\deg L_{2q}^0=2q$;

$\bullet$ a skew-symmetric operation $[\cdot,\cdot]$ such that
\begin{align*}
[L_{2q_1}^0,L_{2q_2}^0] &= 2(q_2-q_1)L_{2(q_1+q_2)}^0, \\
[L_{2q_1},v_{2k}L_{2q_2}] &= v_{2q_1,2k}L_{2q_2}+v_{2k}[L_{2q_1},L_{2q_2}], \\
[v_{2k_1}L_{2q_1},v_{2k_2}L_{2q_2}] &= v_{2k_1}v_{2q_1,2k_2}L_{2q_2} - 
v_{2k_2}[L_{2q_2},v_{2k_1}L_{2q_1}],
\end{align*}
where $v_{2q,2k}\in A(N)$ is a homogeneous polynomial 
$v_{2q,2k}(v_2,\dots,v_{2N})$
of degree $2(q+k)$.
\end{Def}

Using the identity $v_{k_1}(v_{k_2}L_{2q})=(v_{k_1}v_{k_2})L_{2q}$ and Leibniz'
rule, we see that the skew-symmetric operation
$[\cdot,\cdot]$ on the Lie algebra $W_{-1}(N)$
is completely determined by the set of homogeneous polynomials
$v_{2q,2k}\,{=}\,v_{2q,2k}(v_2,\dots,v_{2N})$.

\begin{The}\label{t-3}
The set of polynomials $v_{2q,2k}=v_{2q,2k}(v_2,\dots,v_{2N})\,{\in}\,A(N)$
determines a skew-symmetric operation on an $N$\!-polynomial
Lie algebra $W_{\!-1}(N)$
if and only if the homomorphism
$$
\gamma \colon W_{-1}(N) \to \mbox{Der} A(N),\qquad\gamma(L_{2q}^0) = 
\sum_{k=1}^N v_{2q,2k}\frac{\partial}{\partial v_{2k}},
$$
of $A(N)$-modules is
a homomorphism of the $N$-polynomial Lie algebra $W_{-1}(N)$
to the Lie algebra of polynomial derivations of the ring
$A(N)=\mathbb{C}[v_2,\dots,v_{2N}]$.
\end{The}

{\bf Poof} 
The theorem is proved by a direct verification
of its statements.\\
\hfill $\Box $

The Lie algebra $W_{-1}$ with generators $\mathcal{L}_{2q}^0$, $q=-1,0,1,\dots$,
contains the Lie subalgebra generated by the three operators
$\mathcal{L}_{-2}^0$, $\mathcal{L}_{0}^0$, and $\mathcal{L}_{2}^0$,
where $[\mathcal{L}_{-2}^0,\mathcal{L}_{2}^0]=4\mathcal{L}_{0}^0$.
The Lie algebra $W_{-1}$ with respect to the bracket $[\cdot,\cdot]$
is generated by only two generators, $\mathcal{L}_{-2}^0$ and 
$\mathcal{L}_{4}^0$.

\begin{Ex}
$6\mathcal{L}_{2}^0=[\mathcal{L}_{-2}^0,\mathcal{L}_{4}^0]$,
$4\mathcal{L}_{0}^0=[\mathcal{L}_{-2}^0,\mathcal{L}_{2}^0]$,
and $2\mathcal{L}_{6}^0=[\mathcal{L}_{2}^0,\mathcal{L}_{4}^0]$.
\end{Ex}

The generators $\mathcal{L}_{2q}$, where $q\ge1$, are given by the
recurrence relation
$2q\mathcal{L}_{2(q+2)}^0 = [\mathcal{L}_{2}^0,\mathcal{L}_{2(q+1)}^0]$.
Moreover, the operators $\mathcal{L}_{-2}^0$, $\mathcal{L}_{4}^0$ are related
by commutation relations,
the first of which is
\begin{equation}\label{ex-5}
[\mathcal{L}_{2}^0,[\mathcal{L}_{2}^0,[\mathcal{L}_{2}^0,\mathcal{L}_{4}^0]]]= 
12[\mathcal{L}_{4}^0,[\mathcal{L}_{2}^0,\mathcal{L}_{4}^0]].
\end{equation}

\begin{Cor}\label{cor-8}
The representations $\gamma^j(\mathcal{L}_{2q}^0)=\sum_{k=1}^N 
v_{2q,2k}^j\frac{\partial}{ \partial v_{2k}}$, $j=1,2$,
of the $N$--polynomial algebra $W_{-1}$ coincide if and only if
$v_{2q,2k}^1\equiv v_{2q,2k}^2$ for $q=-1$ and $2$.
\end{Cor}

By construction there is an embedding of the Lie algebra $W_{-1}$ into the Lie 
algebra $W_{-1}(N)$.
On the other hand, the ring homomorphism $\varphi \colon A(N) \to \mathbb{C}$, 
$\varphi(v_{2k})=0$, $k=1,\dots,N$,
induces a projection $W_{-1}(N) \to W_{-1}$ of Lie algebras.

\begin{Cor}\label{cor-9}
The homomorphism
$$
\gamma \colon W_{-1}(N) \to \mathcal{G}_{P,0}(N),\qquad\gamma(L_{2q}^0) = 
\mathcal{L}_{2q}^0 = \sum_{k=1}^N 
2k\mathcal{N}_{2(q+k)}\frac{\partial}{\partial 
\mathcal{N}_{2k}}, \quad \gamma(v_{2k}) = \mathcal{N}_{2k},
$$
extends to an epimorphism of Lie algebras.
\end{Cor}

Note that the nontrivial relation (\ref{f-13})
between Newton polynomials in $x_1,\dots,x_{N}$ ensures the
fulfillment of the condition
$$
\gamma([L_{2k},L_{2k}]) = [\gamma(L_{2k}),\gamma(L_{2k})].
$$
The kernel of the homomorphism $\gamma$ is described by (\ref{f-11}).
The restriction of the homomorphism $\gamma$ to the Lie subalgebra $W_{-1}$
gives a representation of the Lie algebra $W_{-1}$ in the Lie algebra 
$\mathcal{G}_{P,0}(N)$
with the structure of a free $N$-dimensional 
$\mathbb{C}[\mathcal{N}_2,\dots,\mathcal{N}_{2N}]$-module.

\section{Commuting Vector Fields on the Symmetric Square of a Plane Curve}

Consider the symmetric square of the curve
$V\,{=}\,\{ (X,Y)\,{\in}\,\mathbb{C}^2:F(X,Y)\,{=}\,0 \}$,
where $F(X,Y)$ are polynomials in $X$ and $Y$.
Let 
$\mathcal{D}_k=F(X_k,Y_k)_{Y_k}\partial_{X_k}-F(X_k,Y_k)_{X_k}\partial_{Y_k}$, 
$k=1,2$.
We introduce the operators
\begin{equation}\label{f6-2}
\mathcal{L}^1=\frac{1}{X_1-X_2}\,(\mathcal{D}_1-\mathcal{D}_2), \quad 
\mathcal{L}^2=\frac{1}{X_1-X_2}\,(X_2\mathcal{D}_1-X_1\mathcal{D}_2).
\end{equation}

\begin{Lem}\label{L-7}\;\;
1. The operators $\mathcal{L}^1$ and $\mathcal{L}^2$ are derivations of the 
function ring on
$\mbox{Sym}^2(\mathbb{C}^2)\setminus\{X_1-X_2=0\}$.

2. The operators $\mathcal{L}^1$ and $\mathcal{L}^2$ annihilate the polynomials
$F(X_1,Y_1)$ and $F(X_2,Y_2)$.

3. $[\mathcal{L}^1,\mathcal{L}^2]\equiv 0$.
\end{Lem}

{\bf Poof} 
The operators $\mathcal{L}^1$ and $\mathcal{L}^2$ are derivations of the 
function ring on
$(\mathbb{C}^2\times\mathbb{C}^2)\setminus\{X_1-X_2\break=0\}$.
Statements~1 and~2 are verified directly. A standard calculation shows that
$$
[(X_1-X_2)\mathcal{L}^1,(X_1-X_2)\mathcal{L}^2]
=-F(X_2,Y_2)_{Y_2}\mathcal{D}_1-F(X_1,Y_1)_{Y_1}\mathcal{D}_2 + 
(X_1-X_2)^2[\mathcal{L}^1,\mathcal{L}^2].
$$
On the other hand,
$[\mathcal{D}_1-\mathcal{D}_2,X_2\mathcal{D}_1-X_1\mathcal{D}_2] = 
-F(X_2,Y_2)_{Y_2}\mathcal{D}_1-F(X_1,Y_1)_{Y_1}\mathcal{D}_2$.
The coincidence of the left-hand sides of the equations
and the relation $X_1-X_2\neq 0$ imply the lemma.\\
\hfill $\Box $

\section{Lie Algebroids on the Space of Nonsingular Hyperelliptic Curves}

Consider the bundle $f \colon \mathcal{E}_{N,0}\to \mathcal{B}_{N,0}$
(see Definition~\ref{def-1}).
In Section~2 we described the Lie algebra of vector fields
on $\mathcal{B}_{N,0}$
generated by the Newton fields $\mathcal{L}_{2k}^0$, $k=-1,0,1,\dots,N-2$.
In this section we construct a Lie algebroid on the space $\mathcal{E}_{N,0}$.
We set $\pi=\pi(X,Y;\bx)=Y^2-P$, where $P=P(X;\bx)=\prod_{i=1}^N(X-x_i)$.
By $\mathcal{G}(\mathbb{C}[X,Y;\bx])$ we denote
the Lie algebra of derivations of the ring $\mathbb{C}[X,Y;\bx]$.
Let us introduce the operator
$\mathcal{L}_{N-2}^*=2Y\partial_X+P_X\partial_Y \in 
\mathcal{G}(\mathbb{C}[X,Y;\bx])$.
We have $\mathcal{L}_{N-2}^*\pi\equiv 0$.
Hence, for fixed $\bx$, the operator $\mathcal{L}_{N-2}^*$
determines a vector field on $\mathbb{C}^2$
that is tangent to the curve $V=\{ (X,Y)\in \mathbb{C}^2:\pi(X,Y;\bx)=0 \}$.
The field $\mathcal{L}_{N-2}^*$ determines the vertical field of
the bundle $f \colon \mathcal{E}_{N,0}\to \mathcal{B}_{N,0}$.

\begin{Lem}\label{L-8}
Let $\mathcal{D}$ be a derivation of the form $a\partial_X+b\partial_Y$
of the ring $\mathbb{C}[X,Y;\bx]$, where $\bx \in \mathcal{B}_{N,0}$.
Then $\mathcal{D}\pi=\Phi\pi$, where $\Phi\in \mathbb{C}[X,Y;\bx]$,
implies $\mathcal{D}=\psi\mathcal{L}_{N-2}^*+\pi\mathcal{D}^1$,
where $\psi\in\mathbb{C}[X,Y;\bx]$ and 
$\mathcal{D}^1\in\mathcal{G}(\mathbb{C}[X,Y;\bx])$.
\end{Lem}

{\bf Poof} 
We shall carry out calculations in the ring 
$K\,{=}\,\mathbb{C}[X,Y;\bx]/\langle\pi\rangle$.
In this ring $Y^2=P$, and thus $K$ is a free $\mathbb{C}[X;\bx]$-module with 
generators $1$ and $Y$.
We set $a=a_1+a_2Y$ and $b=b_1+b_2Y$, where $a_l,b_l\in\mathbb{C}[X;\bx]$, 
$l=1,2$.
The condition that $\mathcal{D}\pi=0$ in the ring $K$
implies $(a_1+a_2Y)P_X=(b_1+b_2Y)2Y$.
Hence $a_1P_X=2b_2P$ and $a_2P_X=2b_1$. On the other hand,
the condition $\mathcal{D}=\psi\mathcal{L}_{N-2}^*$,
where $\psi=\psi_1+\psi_2Y$, implies
$$
a_1+a_2Y=2\psi_2P+2\psi_1Y, \qquad b_1+b_2Y=\psi_1P_X+\psi_2P_XY.
$$
Hence $2\psi_1=a_2$, $2\psi_2P=a_1$, and $\psi_2P_X=b_2$.
Since $\bx \in \mathcal{B}_{N,0}$, it follows that
the polynomials $P(X;\bx)$ and $P_X(X;\bx)$
are coprime, and this system has a polynomial solution 
$\psi_2=\psi_2(X;\bx)$.\\
\hfill $\Box $

Consider the following sequence of derivations of the ring 
$\mathbb{C}[X,Y;\bx]$:
\begin{equation}\label{diff}
{L}_{2k}^0=\mathcal{L}_{2k}^0 + 2X^{k+1}\partial_X + 
C_{2k}Y\partial_Y,\quad\text{where }\,C_{2k}=\sum_{i=1}^N 
\frac{X^{k+1}-x_i^{k+1}}{X-x_i}\,.
\end{equation}

\begin{The}\label{t-4}
The homogeneous fields ${L}_{2k}^0$, $k=-1,0,1,\dots$, of degree $2k$
are uniquely determined by
the condition that they are lifts
of the Newton fields $\mathcal{L}_{2k}^0$ and generate the Lie algebra
of Newton horizontal vector fields on the space of the bundle 
$\mathcal{E}_{N,0}$, that is,
$$
[{L}_{2q_1}^0,{L}_{2q_2}^0]=2(q_2-q_1){L}_{2(q_1+q_2)}^0.
$$
\end{The}

{\bf Poof} 
We set 
$\widehat{\mathcal{L}}_{2k}^0=\mathcal{L}_{2k}^0+2X^{k+1}\!\partial_X=2(\sum_{
i=1}^N x_i^{k+1}\!\partial_{x_i} + X^{k+1}\!\partial_X)$.
The operator $\widehat{\mathcal{L}}_{2k}^0$ determines a Newton derivation of 
the ring $\mathbb{C}[X;\bx]$.
It is easy to check that (\ref{diff}) can be written as
\begin{equation}\label{f-28}
{L}_{2k}^0=\widehat{\mathcal{L}}_{2k}^0+\frac{1}{2}(\widehat{\mathcal{L}}_{2k}
^0\ln P)Y\partial_Y.
\end{equation}
Hence
${L}_{2k}^0(Y^2-P)=P(\widehat{\mathcal{L}}_{2k}^0\ln 
P-\widehat{\mathcal{L}}_{2k}^0\ln P)\equiv 0$.
Thus, formula (\ref{diff}) determines horizontal vector fields ${L}_{2k}^0$, 
$k=-1,0,1,\dots$,
on $\mathcal{E}_{N,0}$, which are lifts of the fields $\mathcal{L}_{2k}^0$ on 
the base $\mathcal{B}_{N,0}$.

Now let ${L}_{2k}^{0,1}$ and ${L}_{2k}^{0,2}$ be two homogeneous horizontal 
vector fields
on $\mathcal{E}_{N,0}$ that are
lifts of the field $\mathcal{L}_{2k}^0$ on the base $\mathcal{B}_{N,0}$.
Then, according to Lemma~\ref{L-8},
${L}_{2k}^{0,2}={L}_{2k}^{0,1} + \psi_{2k+2-N}\mathcal{L}_{N-2}^*$,
where $\psi_{2k+2-N}=\psi_1+\psi_2Y$ and $\psi_1,\,\psi_2 \in \mathbb{C}[X;\bx]$
are homogeneous polynomials such that $\deg \psi_1=2m=2k+2-N$ and $\deg 
\psi_2=2(k+1-N)$.

Note that the degree of the function $\psi_{2k+2-N}$ cannot be negative.
Hence the condition $\psi_{2k+2-N}\break\neq 0$ implies $N\le 2k+2$.
On the other hand, according to Corollary~\ref{cor-6}, the generators of the 
algebra $W_{-1}$
are completely determined by the operators ${L}_{-2}^0$ and ${L}_{4}^0$.
As a result, we obtain the following conditions: $N\le 0$ for $k=-1$,
$N\le 4$ for $k=1$, and $N\le 6$ for $k=2$.
In the case where $k=2$ and $N=5$, we obtain
$\deg \psi_1=1$, which contradicts $\deg \psi_1=2m$.
Thus, we have proved that the lift of the fields $\mathcal{L}_{2k}^0$, 
$k=-1,0,1,\dots$,
is unique for $N=5$ and $N>6$.

It remains to consider the cases $N=3,4,6$.
As shown above, in the case $N=6$, the lifts of the fields
$\mathcal{L}_{-2}^0$ and $\mathcal{L}_{2}^0$
are unique, and any lift of $\mathcal{L}_{4}^0$ must
have the form $L_4^0+\alpha\mathcal{L}_{4}^*$, where $\alpha\in\mathbb{C}$.
A direct verification shows that the commutation relation (see \eqref{ex-5})
in the Witt algebra holds only for $\alpha=0$.
Thus, in the case $N=6$, the lift of the fields $\mathcal{L}_{2k}^0$ is unique.
Similar arguments show that this is also true in the cases $N=3$ and $4$.

The commutation rule $[{L}_{2q_1}^0,{L}_{2q_2}^0]=2(q_2-q_1){L}_{2(q_1+q_2)}^0$
follows from the fact that $\widehat{\mathcal{L}}_{2k}^0$ is a Newton operator
and from (\ref{f-28}).
This completes the proof of the theorem.\\
\hfill $\Box $

\begin{Cor}\label{cor-10}
The generating function for the operators (\ref{diff}) has the form
\begin{equation}\label{pr}
L^0(t)=\widehat{\mathcal{L}}^0(t)+\frac{1}{2}(\widehat{\mathcal{L}}^0(t)\ln 
P)Y\partial Y,\quad\text{where 
}\,\widehat{\mathcal{L}}^0(t)=\mathcal{L}^0(t)+2\,\frac{1}{1-Xt}\, \partial_X.
\end{equation}
\end{Cor}

Consider the space $\mathbb{C}^{N+1}$ with the graded coordinates
$(X,Y;\mathcal{N}_2,\dots,\mathcal{N}_{2(N-1)})$.
Using the equation $Y^2=P(X;\bx)$, we can identify the space
$\mathcal{E}_{N,0}$ with an open dense subvariety
in $\mathbb{C}^{N+1}$.
The Lie algebra of vector fields on $\mathcal{E}_{N,0}$ described above
determines a polynomial Lie algebra
generated by the field $\mathcal{L}_{N-2}^*$
and the fields $L_{-2}^0,L_0^0,L_2^0,\dots,L_{2(N-2)}^0$.

\begin{Ex} \label{ex-6}
Case $N=3$. The coordinates in $\mathbb{C}^4$ are $X$, $Y$, $\mathcal{N}_2$, and
$\mathcal{N}_4$. We have
$$
\frac{1}{3}\mathcal{N}_6=-Y^2+X^3-\mathcal{N}_2X^2+\frac{1}{2}(\mathcal{N}
_2^2-\mathcal{N}_4)X+ 
\bigg(\frac{1}{2}\mathcal{N}_2\mathcal{N}_4-\frac{1}{6}\mathcal{N}_2^3\bigg).
$$
Using this formula, we obtain an explicit expression for
the basis polynomial fields $\mathcal{L}_1^*$, $L_{-2}^0$, $L_0^0$, and $L_2^0$
in $\mathbb{C}^4$.
\end{Ex}

\section{Coordinate Rings of Spaces of Symmetric Squares
of Hyperelliptic Curves}

Consider the space $\bbbC^2\times\bbbC^2$ with coordinates
$(X_1,Y_1)$ and $(X_2,Y_2)$ graded as above, i.e., so that
$\deg X_k=2$ and $\deg Y_k=N$, $k=1,2$, and the space $\bbbC^5$
with graded coordinates $u_2$, $u_4$, $v_{N}$, $v_{N+2}$, and $v_{2N}$.
Here the subscript corresponds to the degree of variables.

\begin{Lem}\label{L-9}
The algebraic homogeneous map
$$
\xi \colon \bbbC^2\times\bbbC^2 \to \bbbC^5,\qquad 
\xi((X_1,Y_1),(X_2,Y_2))=(u_2,u_4,v_{N},v_{N+2},v_{2N}),
$$
where
$u_2=X_1+X_2$, $u_4=(X_1-X_2)^2$, $v_{N}=Y_1+Y_2$, $v_{N+2}=(X_1-X_2)(Y_1-Y_2)$,
and $v_{2N}=(Y_1-Y_2)^2$,
makes it possible to identify the algebraic variety $(\bbbC^2\times\bbbC^2)/S_2$
with the hypersurface in $\bbbC^5$
determined by the equation $u_4v_{2N}-v_{N+2}^2=0$.
\end{Lem}

{\bf Poof} 
The lemma is proved directly.\\
\hfill $\Box $

For what follows we need the homogeneous polynomials
$a_{2k}(u_2,u_4)$ of degree $2k$
determined by the generating series
\begin{align}
\frac{1}{(1-X_1t)(1-X_2t)}&= a(t;u_2,u_4)=\sum_{k=0}^\infty 
a_{2k}(u_2,u_4)t^k\notag\\
&=4\,\frac{1}{(2-u_2t)^2-u_4t^2}=1+u_2t+(t^2).\label{f-32}
\end{align}

In the notation of Lemma~\ref{L-9} we have
\begin{equation}\label{f-33}
\sum_{k=0}^\infty (X_1^k+X_2^k)t^k=\frac{1}{1-X_1t}+\frac{1}{1-X_2t}=
(2-u_2t)a(t;u_2,u_4).
\end{equation}
Moreover,
\begin{align}
&\sum_{k=2}^\infty(X_1-X_2)(X_1^{k-1}-X_2^{k-1})t^{k-2}\notag\\
&\qquad=\sum_{k=2}^\infty [(X_1^k+X_2^k)t^{k-2}-
X_1X_2(X_1^{k-2}+X_2^{k-2})t^{k-2}]=u_4a(t;u_2,u_4),\label{f-34}\\
\label{f-35}
&\sum_{k=0}^\infty (Y_1-Y_2)(X_1^k-X_2^k)t^{k} = 
(Y_1-Y_2)\bigg[\frac{1}{1-X_1t}-\frac{1}{1-X_2t}\bigg]=
v_{N+2}ta(t;u_2,u_4).
\end{align}
We have $Y_j^2=X_j^{N} + \sum_{k=2}^N (-1)^k y_{2k}X_j^{N-k}$. Hence
\begin{align*}
(Y_1^2+Y_2^2)&=\frac{1}{2}(v_{N}^2+v_{2N})\\
&=(2a_{2N}-u_2a_{2N-2})
+\sum_{k=2}^{N-1} (-1)^k y_{2k}(2a_{2(N-k)}-u_2a_{2(N-k-1)}) +(-1)^N 2y_{2N}.
\end{align*}
We also have
\begin{align*}
(X_1-X_2)(Y_1^2-Y_2^2) &= v_{N}v_{N+2} = u_4\bigg(a_{2(N-2)}+ \sum_{k=2}^{N-2} 
(-1)^k y_{2k}a_{2(N-k-2)}\bigg),\\
(Y_1-Y_2)(Y_1^2-Y_2^2) &= v_{N}v_{2N} = 
v_{N+2}\bigg(a_{2(N-1)}+\sum_{k=2}^{N-1} (-1)^k y_{2k}a_{2(N-k-1)}\bigg).
\end{align*}

The graded coordinate ring $\mathcal{R}_0(N)$ of the space 
$\widehat{\mathcal{E}}_{N,0}^2$
in 
$(\mathbb{C}^2\times\mathbb{C}^2\setminus\{X_1-X_2=0\})\times\mathcal{B}_{N,0}$
(see Section~1) has the form
\begin{gather*}
\mathbb{C}[X_1,Y_1;X_2,Y_2;x_1,\dots,x_N]/J,\\
\deg x_j=\deg X_k=2,\;\deg Y_k=N,\;j=1,\dots,N,\;k=1,2,
\end{gather*}
where $J=\langle\pi_1,\pi_2\rangle$ is the ideal generated by the polynomials
$\pi_k=\pi_k(X_k,Y_k;\bx)$.
Let $\mathcal{R}_0^G(N)\subset\mathcal{R}_0(N)$ denote
the invariant ring of the free action of $G$ on $\mathcal{R}_0(N)$.
Consider the graded
ring $\mathcal{R}(N)=\mathcal{R}_0(N)/\langle y\rangle$, where 
$y=x_1+\cdots+x_N$.
Let $\mathcal{R}^G(N)\subset \mathcal{R}(N)$ denote
the invariant ring of the free action of $G$ on $\mathcal{R}(N)$.
We shall treat the ring $\mathcal{R}^G(N)$
as the coordinate ring of the universal space $\mathcal{E}_{N}^2$.

\begin{Lem}\label{L-10}
The ring $\mathcal{R}^G(N)$ is isomorphic to the graded ring
$$
\cR_U^G=\mathbb{C}[u_2,u_4,v_{N},v_{N+2},v_{2N},{\bf y}]/ J^G,
$$
where ${\bf y}=(y_4,\dots,y_{2N})$ and the ideal $J^G$ has
Gr\"obner basis
\begin{align*}
P_{2N+4} &= v_{N+2}^2-u_4v_{2N},\\
P_{2N+2} &= v_{N}v_{N+2}-u_4\bigg(a_{2N-2}+\sum_{k=2}^{N-1} (-1)^k 
y_{2k}a_{2(N-k-1)}\bigg),\\
P_{2N} &= v_{N}^2+v_{2N}-(a_{2N}-u_2a_{2(N-1)})
-\sum_{k=2}^{N-1} (-1)^k y_{2k}(2a_{2(N-k)}-u_2a_{2(N-k-1)})- (-1)^N 
2y_{2N},\\[-7pt]
P_{3N} &= v_{N}v_{2N}-v_{N+2}\bigg(a_{2(N-1)}+\sum_{k=2}^{N-1} (-1)^k 
y_{2k}a_{2(N-k-1)}\bigg).
\end{align*}
The relation $v_{N}P_{2N+4}-v_{N+2}P_{2N+2}+u_4P_{3N}=0$ holds.
\end{Lem}

{\bf Poof} 
The lemma follows easily from the relations
obtained above.\\
\hfill $\Box $

Let us introduce the ring 
$\mathcal{A}(N)=\mathbb{C}[u_2,u_4,v_{N-2},v_{N},\widehat{{\bf y}}]$,
where $\widehat{{\bf y}}=(y_4,\dots,y_{2(N-2)})$.

\begin{Lem}\label{L-11}
There is a ring homomorphism $\varphi\colon\cR_U^G\to \mathcal{A}(N)$ defined 
by:
\begin{eqnarray}
 &&\varphi(u_{2})=u_{2}, \quad \varphi(u_{4})=u_{4},\quad 
\varphi(v_N)=v_N,\quad\varphi(y_{2k})=y_{2k},\qquad k=2,\dots,N-2,\\ \nonumber 
&&\\
&& \varphi(v_{N+2})=u_4 v_{N-2},\quad\varphi(v_{2N})=u_4 v_{N-2}^2,\\  
\label{phiy2N2}
&&\varphi(y_{2(N-1)}) =(-1)^{N-1} 
\bigg[v_{N-2}v_{N}-a_{2(N-1)}-\sum_{k=2}^{N-2} 
(-1)^k y_{2k}a_{2(N-k-1)}\bigg],\\ \label{phiy2N}
&&\varphi(y_{2N}) = 
(-1)^{N}\,\frac{1}{2}\bigg[(v_{N}^2+v_{2N})-(2a_{2N}-u_2a_{2(N-1)}) \nonumber 
\\ 
&&\kern60pt-\sum_{k=2}^{N-1} (-1)^k 
y_{2k}(2a_{2(N-k)}-u_2a_{2(N-k-1)})\bigg]. 
\end{eqnarray}

\end{Lem}

{\bf Poof} 
A direct verification shows that the homomorphism $\varphi$ maps the ideal 
$J^G$ to $0$.\\
\hfill $\Box $

\begin{Cor}\label{cor-11}
The homomorphism $\varphi[u_4^{-1}] \colon \cR_U^G[u_4^{-1}] \to 
\mathcal{A}(N)[u_4^{-1}]$
is an isomorphism.
\end{Cor}

{\bf Poof} 
The ring homomorphism
\begin{gather*}
\eta \colon \mathcal{A}(N)[u_4^{-1}] \to \cR_U^G[u_4^{-1}],\\
\eta(u_{2k})=u_{2k},\quad 
k=1,2,\qquad\eta(v_N)=v_N,\quad\eta(y_{2k})=y_{2k},\quad k=2,\dots,N-2,\\
\eta(v_{N-2})=u_4^{-1}v_{N+2}
\end{gather*}
is inverse to the homomorphism $\varphi[u_4^{-1}]$.\\
\hfill $\Box $

Consider the space $\mathbb{C}^{N+4}$ with the graded coordinates
$(u_2,u_4,v_{N},v_{N+2},v_{2N};{\bf y})$
and the space $\mathbb{C}^{N+1}$ with the graded coordinates
$(u_2,u_4,v_{N-2},v_{N},\widehat{{\bf y}})$.
As mentioned above, the space $\mathcal{E}^2_N$
can be identified with the algebraic subvariety in $\mathbb{C}^{N+4}$
determined by the equations $P_{2N+k}= 0$, $k=0,2,4,N$ (see Lemma~\ref{L-10}).
We set
$$
b_{2N}(u_2;\widehat{{\bf y}})=\frac{1}{2^{N-1}}\bigg(u_2^N+ \sum_{k=2}^{N-1} 
(-1)^k 2^k y_{2k}u_2^{N-k} +
(-1)^N 2^N y_{2N}\bigg).
$$
Let $W_s(N)$, $s=1,2$, denote the algebraic subvarieties
in $\mathbb{C}^{N+4}$ determined by the equations
\begin{eqnarray*}
&&\text{for }\,s=1:\quad u_4=0,\quad v_N=0,\quad v_{N+2}=0,\quad 
v_{2N}=b_{2N}(u_2;\widehat{{\bf y}},\\
&&\text{for }\,s=2:\quad u_4=0,\quad v_{N+2}=0,\quad v_{2N}=0,\quad 
v_{N}^2=b_{2N}(u_2;\widehat{{\bf y}}.
\end{eqnarray*}

We set $W(N)=W_1(N)\cup W_2(N)$.
Note that, for given ${\bf y}$,
the intersection $W_1(N)\cap W_2(N)$ is the set of roots of the equation
$b_{2N}(u_2;\widehat{{\bf y}}=0$.
Clearly, $W(N)\subset \mathcal{E}^2_N$.

\begin{The}\label{t-5}
The mapping
$f \colon \mathbb{C}^{N+1}\to\mathbb{C}^{N+4}$ defined by
$f(u_2,u_4,v_{N-2},v_{N},\widehat{{\bf y}})= (u_2,u_4,v_{N},\break 
v_{N+2},v_{2N};{\bf y})$,
where $v_{N+2}=u_4v_{N-2}$, $v_{2N}=u_4v_{N-2}^2$,
\begin{align*}
y_{2(N-1)} &=(-1)^{N-1}\bigg[ v_{N}v_{N-2}-\bigg(a_{2(N-1)} +
\sum_{k=2}^{N-2} (-1)^k y_{2k}a_{2(N-k-1)} \bigg)\bigg], \\
\intertext{and}
y_{2N} &=(-1)^{N}\frac{1}{2}\bigg[ v_{N}^2+v_{2N} - (2a_{2N}-u_2a_{2(k-1)})
-\sum_{k=2}^{N-1} (-1)^k y_{2k}(2a_{2(N-k)}-u_2a_{2(N-k-1)})\bigg],
\end{align*}
determines a homomorphism $f \colon \mathbb{C}^{N+1}\setminus\{u_4=0\}\to 
\mathcal{E}^2_N\setminus W(N)$.
\end{The}

{\bf Poof} 
The required assertion follows directly from Lemmas~\ref{L-10} and~\ref{L-11}
and Corollary~\ref{cor-11}.\\
\hfill $\Box $

\section{Lie Algebroids on the Space of Symmetric Squares\\
of Nonsingular Hyperelliptic Curves}

Consider the bundle $\mathcal{E}_{N,0}^2\to \mathcal{B}_{N,0}$
(see Definition~\ref{def-2}).
We have defined an action
of the Witt algebra of Newton fields $\mathcal{L}_{2q}^0$
(see Definition~\ref{def-3}) on the base $\mathcal{B}_{N,0}$.
Let us introduce the following derivations of the ring 
$\mathbb{C}[X_1,Y_1;X_2,Y_2;\bx]$:
\begin{equation}\label{f-36}
{L}_{2k}^0=\mathcal{L}_{2k}^0 + 
2(X_1^{k+1}\partial_{X_1}+X_2^{k+1}\partial_{X_2}) + C_{2k}^1Y_1\partial_{Y_1} 
+ C_{2k}^2Y_2\partial_{Y_2},
\end{equation}
where
$$
C_{2k}^j=\sum_{i=1}^N \frac{X_j^{k+1}-x_i^{k+1}}{X_j-x_i},\qquad j=1,2.
$$
We set 
$\widehat{\mathcal{L}}_{2k}^0=\mathcal{L}_{2k}^0+2(X_1^{k+1}\partial_{X_1}+X_2^{
k+1}\partial_{X_2})$.
It is verified directly that
\begin{equation}\label{f-38}
{L}_{2k}^0=\widehat{\mathcal{L}}_{2k}^0+\frac{1}{2}\widehat{\mathcal{L}}_{2k}^0[
(\ln P^{(1)}) Y_1\partial_ {Y_1} + (\ln P^{(2)}) Y_2\partial_{Y_2}], \quad 
\text{where }\,P^{(j)}=\prod_{i=1}^N(X_j-x_i).
\end{equation}
Now we are ready to obtain one of the main results of the
present paper.

\begin{The}\label{t-6}
The homogeneous fields ${L}_{2k}^0$, $k=-1,0,1,\dots$, of degree $2k$
(see (\ref{f-36})) are determined uniquely by the conditions
that they are lifts of the Newton fields $\mathcal{L}_{2k}^0$
on the base of the bundle $\mathcal{E}_{N,0}^2\to \mathcal{B}_{N,0}$,
generate the Lie algebra of Newton horizontal vector fields on 
$\mathcal{E}_{N,0}^2$,
and determine a representation of the Lie algebra $W_{-1}$.
\end{The}

{\bf Poof} 
The proof of the theorem uses explicit expressions
and Lemma~\ref{L-8} by analogy with the proof of Theorem~\ref{t-4}.\\
\hfill $\Box $

Using the operators $\mathcal{D}_k=2Y_k\partial_{X_k}+P_{X_k}^{(k)}\partial{ 
Y_k}$, $k=1,2$,
we infer (see Lemma~9) that in the Lie algebroid
of the bundle $\mathcal{E}_{N,0}^2$
there are the two commuting horizontal fields
\begin{equation}\label{2-kom}
\mathcal{L}_{N-4}^* =\frac{1}{X_1-X_2}\,(\mathcal{D}_1-\mathcal{D}_2)
 \qquad\mbox{and}\qquad
\mathcal{L}_{N-2}^* =\frac{1}{X_1-X_2}\,(X_2\mathcal{D}_1-X_1\mathcal{D}_2).
\end{equation}

\begin{Lem}\label{LL-14}
For the curve $Y^2 = \prod_{i=1}^N (X-x_i)$,
$$
E(t)-(1-tX)\sum_{m=1}^N E(t;m)X^{m-1} t^{m-1} = t^NY^2.
$$
\end{Lem}

{\bf Poof} 
We have
\begin{align*}
\quad&E(t)-\sum_{m=1}^N E(t;m)X^{m-1} t^{m-1}+\sum_{m=1}^N E(t;m)X^{m} t^{m}\\
&\qquad= (E(t)-E(t;1))+\sum_{m=1}^{N-1}(E(t;m)-E(t;m+1))X^{m} t^{m}+ 
E(t;N)X^{N} t^{N}\\
&\qquad= (-1)^Ny_{2N}t^{N} + \sum_{m=1}^{N-1}(-1)^m y_{2m}X^{m} t^{N}+X^{N} 
t^{N}= t^NY^2.\kern70pt\Box
\end{align*}
\hfill $\Box $

Let $L(t)=\sum_{k=-1}^\infty L_{2k}^0t^{k+1}$.

\begin{The}\label{t-7}
For the generating series $L(t)$ of the operators
$L_{2k}^0$, $k=-1,0,1,\dots$ (see (\ref{f-36})),
the relation
\begin{equation}\label{F-36}
E(t)L(t)=\sum_{m=1}^N E(t;m) 
t^{m-1}{L}_{2(m-2)}^0+\mathcal{A}_2(t)\mathcal{L}_{N-4}^* - 
\mathcal{A}_0(t)\mathcal{L}_{N-2}^*
\end{equation}
holds on the variety $(X_1-X_2\neq 0,Y_1Y_2\neq 0)$, where
\begin{align*}
\mathcal{A}_2(t)&=t^{N}\bigg[ Y_1\frac{X_1}{1-tX_1}+Y_2\frac{X_2}{1-tX_2}\bigg]
= t^{N}a(t;u_2,u_4)\bigg[ \frac{1}{2}(u_2v_N+v_{N+2})- 
\frac{1}{4}t(u_2^2-u_4)\bigg],\\
\mathcal{A}_0(t)&=t^{N}\bigg[ Y_1\frac{1}{1-tX_1}+Y_2\frac{1}{1-tX_2}\bigg] = 
t^{N}a(t;u_2,u_4)\bigg[ v_N-\frac{1}{2}t(u_2v_N-v_{N+2})\bigg].
\end{align*}
\end{The}

{\bf Poof} 
Let $\mathfrak{L}=E(t)L(t)-\sum_{m=1}^N E(t;m) t^{m-1}{L}_{2(m-2)}^0$.
According to Corollary~\ref{cor-3}, we have $\mathfrak{L}x_i=0$, $i=1,\dots,N$.
Thus, the field $\mathfrak{L}$ is vertical, and therefore
$\mathfrak{L}=\mathcal{A}_2(t)\mathcal{L}_{N-4}^* - 
\mathcal{A}_0(t)\mathcal{L}_{N-2}^*$
for some series $\mathcal{A}_2(t)$ and $\mathcal{A}_0(t)$.
On the other hand, according to Lemma~\ref{LL-14},
$$
\mathfrak{L}X_j=\frac{1}{1-tX_j}\bigg[ E(t)-(1-tX_j)\sum_{m=1}^N 
E(t;m)X_j^{m-1} t^{m-1}\bigg] = \frac{t^NY_j^2}{1-tX_j}\,.
$$
Using (\ref{2-kom}), we obtain
the system of equations
$$
\frac{1}{X_1-X_2}(\mathcal{A}_2(t)-X_2\mathcal{A}_0(t)) = 
\frac{t^NY_1}{1-tX_1},\qquad
\frac{1}{X_2-X_1}(\mathcal{A}_2(t)-X_1\mathcal{A}_0(t)) = 
\frac{t^NY_2}{1-tX_2}\,,
$$
provided that $Y_1Y_2\neq 0$.
Solving this system completes the proof of the theorem.\\
\hfill $\Box $

\begin{Cor}\label{cor-12}
In the basis $\{ 
{L}_{-2}^0,\dots,{L}_{2(N-2)}^0;\mathcal{L}_{N-4}^*,\mathcal{L}_{N-2}^* \}$
the following commutation relations hold:
$$
[{L}_{2p},{L}_{2q}]=2(q-p){L}_{2(p+q)}=2(q-p)\bigg(\sum_{m=1}^N 
\omega_{p+q,m}{L}_{2(m-2)}+\alpha_{p+q}\mathcal{L}_{N-4}^*-\beta_{p+q}\mathcal{L
}_{N-2}^*\bigg),
$$
where $\omega_{p+q,m}$, $\alpha_{p+q}$, and $\beta_{p+q}$
are the coefficients of $t^{p+q}$ in the series $E(t;m)/E(t)$,
$\mathcal{A}_2(t)/E(t)$, and $\mathcal{A}_0(t)/E(t)$.
All these coefficients belong to the ring $\mathcal{R}^G(N)$
(see Lemma~\ref{L-10}).
\end{Cor}

We set $\mathcal{N}(t)=\sum_{i=1}^N 1/(1-x_it)$ and
$\mathcal{D}_0(t)=\mathcal{N}(t)-2(1/(1-tX_1)+1/(1-tX_2)) = 
\mathcal{N}(t)-2a(t)(2-u_2t)$,
where $a(t)=a(t;u_2,u_4)$ (see (\ref{f-32})).

\begin{Lem}\label{LL-15}
The following relations hold:
\begin{gather}\label{F-37}
[L(t),\mathcal{L}_{N-4}^*]X_1=\frac{t}{1-tX_1}\mathcal{D}_0(t)\mathcal{L}_{N-4}
^*X_1,\\
\label{F-38}
[L(t),\mathcal{L}_{N-2}^*]X_1=\bigg(\frac{2}{1-tX_2}+\frac{tX_2}{1-tX_1}\mathcal
{D}_0(t) \bigg) \mathcal{L}_{N-4}^* X_1.
\end{gather}
\end{Lem}

{\bf Poof} 
Using (\ref{f-36}) and (\ref{2-kom}), we obtain
\begin{alignat*}2
L(t)X_1 &= \frac{2}{1-tX_1},&\qquad L(t)Y_1&= \frac{tY_1}{1-tX_1} 
\mathcal{N}(t),\\
\mathcal{L}_{N-4}^*X_1 &= \frac{2Y_1}{X_1-X_2},&\qquad \mathcal{L}_{N-2}^*X_1 
&= X_2\mathcal{L}_{N-4}^*X_1.
\end{alignat*}
Hence
\begin{align}
\label{F-39}
L(t)\mathcal{L}_{N-4}^*X_1 &= \frac{t}{1-tX_1} 
\bigg(\mathcal{N}(t)-2\,\frac{1}{1-tX_2} \bigg) \mathcal{L}_{N-4}^*X_1,\\
\label{F-40}
\mathcal{L}_{N-4}^*L(t)X_1 &= \frac{2t}{(1-tX_1)^2}\,\mathcal{L}_{N-4}^* X_1.
\end{align}
Relations (\ref{F-39}) and (\ref{F-40}) imply (\ref{F-37}).
Further, we have $L(t)\mathcal{L}_{N-2}^*X_1 = L(t)(X_2\mathcal{L}_{N-4}^*X_1)$.
Using (\ref{F-39}), we obtain
\begin{align}
\label{F-41}
L(t)\mathcal{L}_{N-2}^*X_1 &= \bigg[\frac{2}{1-tX_2}+\frac{tX_2}{1-tX_1} 
\bigg(\mathcal{N}(t)-2\,\frac{1}{1-tX_2} \bigg)\bigg] \mathcal{L}_{N-4}^*X_1,\\
\label{F-42}
\mathcal{L}_{N-2}^*L(t)X_1 &= \frac{2tX_2}{(1-tX_1)^2}\, \mathcal{L}_{N-4}^* 
X_1.
\end{align}

Relations (\ref{F-41}) and (\ref{F-42}) imply (\ref{F-38}).\\
\hfill $\Box $

We set
\begin{align}
\label{F-45} \mathcal{A}_{-2}(t) &= ta(t)\mathcal{D}_0(t),\quad 
\mathcal{A}_{-4}(t) = t\mathcal{A}_{-2}(t),\\
\label{F-46} \mathcal{B}_{0}(t) &= a(t)\bigg[ 
2(1-u_2t)+\frac{t^2}{4}(u_2^2-u_4)\mathcal{D}_0(t)\bigg],\\
\label{F-47} \mathcal{B}_{-2}(t)&= ta(t)[ 2+(1-u_2t)\mathcal{D}_0(t)].
\end{align}

\begin{The}\label{t-8}
On the variety $\{X_1-X_2\neq 0,\,Y_1Y_2\neq 0\}$
the commutation formulas for the generating series $L(t)$
of horizontal fields with the vertical fields $\mathcal{L}_{N-4}^*$ and
$\mathcal{L}_{N-2}^* $ are
\begin{align}
\label{F-43} [L(t),\mathcal{L}_{N-4}^*] &= 
\mathcal{A}_{-2}(t)\mathcal{L}_{N-4}^* - 
\mathcal{A}_{-4}(t)\mathcal{L}_{N-2}^*,\\
\label{F-44} [L(t),\mathcal{L}_{N-2}^*] &= 
\mathcal{B}_{0}(t)\mathcal{L}_{N-4}^* - \mathcal{B}_{-2}(t)\mathcal{L}_{N-2}^*.
\end{align}
\end{The}

{\bf Poof} 
The series $[L(t),\mathcal{L}_{N-4}^*]$ and $[L(t),\mathcal{L}_{N-2}^*]$
are generating series for the vertical fields.
Let us find their representation in the form of linear combinations of the 
fields
$\mathcal{L}_{N-4}^*$ and $\mathcal{L}_{N-2}^*$.
According to Lemma~\ref{LL-15}, the coefficients $\mathcal{A}_{-2}(t)$, 
$\mathcal{A}_{-4}(t)$,
$\mathcal{B}_{0}(t)$, and $\mathcal{B}_{-2}(t)$
are solutions of the systems of equations
\begin{align*}
\mathcal{A}_{-2}(t)-X_2\mathcal{A}_{-4}(t) &= 
\frac{t}{1-tX_1}\mathcal{D}_0(t),\\
\mathcal{A}_{-2}(t)-X_1\mathcal{A}_{-4}(t) &= \frac{t}{1-tX_2}\mathcal{D}_0(t)
\end{align*}
and
\begin{align*}
\mathcal{B}_{0}(t)-X_2\mathcal{B}_{-2}(t) &= 
\frac{2}{1-tX_2}+\frac{tX_2}{1-tX_1}\mathcal{D}_0(t),\\
\mathcal{B}_{0}(t)-X_1\mathcal{B}_{-2}(t) &= 
\frac{2}{1-tX_1}+\frac{tX_1}{1-tX_2}\mathcal{D}_0(t).
\end{align*}
Solving these systems, we obtain (\ref{F-45})--(\ref{F-47}).
\hfill $\Box $

\begin{Cor}\label{cor-13}
In the basis $\{ 
{L}_{-2}^0,\dots,{L}_{2(N-2)}^0;\mathcal{L}_{N-4}^*,\mathcal{L}_{N-2}^* \}$
the following commutation relations hold:
\begin{align*}
[{L}_{2q},\mathcal{L}_{N-4}^*] &= 
a_{-2,2q+2}\mathcal{L}_{N-4}^*-a_{-4,2q+2}\mathcal{L}_{N-2}^*,\\
[{L}_{2q},\mathcal{L}_{N-2}^*] &= 
b_{0,2q+2}\mathcal{L}_{N-4}^*-b_{-2,2q+2}\mathcal{L}_{N-2}^*,
\end{align*}
where $a_{-2,2q+2}$, $a_{-4,2q+2}$, $b_{0,2q+2}$, and $b_{-2,2q+2}$
are the coefficients of $t^{q+1}$ in the series
$\mathcal{A}_{-2}(t)$, $\mathcal{A}_{-4}(t)$, $\mathcal{B}_{0}(t)$, and 
$\mathcal{B}_{-2}(t)$.
All these coefficients lie in the ring $\mathcal{R}^G(N)$ (see 
Lemma~\ref{L-10}).
\end{Cor}

\section{Polynomial Lie Algebroids Determined by the Lie Algebroid
on $\boldsymbol{\mathcal{E}_{N,0}^2}$}

In this section we give a description of the polynomial Lie algebroid
$\mathcal{G}(N)$ on $\mathbb{C}^{N+1}$, which uses the homomorphism
$f \colon \mathbb{C}^{N+1}\setminus\{u_4=0\}\to\mathcal{E}_{N,0}^2\setminus 
W(N)$
constructed in Section~6.
As generators of the Lie algebroid $\mathcal{G}(N)$ we take
the horizontal vector fields ${L}_{2k}^0$
and the vertical fields $\mathcal{L}_{N-4}^*$ and $\mathcal{L}_{N-2}^*$, which
were constructed in Section~7.
Without loss of generality, it is sufficient
to consider the algebroid $\mathcal{G}$
as a module over the polynomial ring 
$\mathcal{A}(N)=\mathbb{C}[u_2,u_4,v_{N-2},v_N;{\bf y}]$.

\begin{Lem}\label{L-13}
The action of the operators $\mathcal{L}_{N-4}^*$ and $\mathcal{L}_{N-2}^*$ on
the coordinate functions $u_2$ and $u_4$ has the form
\begin{alignat*}2
\cL^{\star}_{N-4}u_2&=2 v_{N-2},&\qquad\cL^{\star}_{N-2}u_2&=u_2 
v_{N-2}-v_{N},\\
\cL^{\star}_{N-4}u_4&=4 
v_{N},&\qquad\cL^{\star}_{N-2}u_4&=2u_2v_{N}-2u_4v_{N-2}.
\end{alignat*}
\end{Lem}

{\bf Poof} 
The required relations are derived directly from our results obtained above.\\
\hfill $\Box $

The action of the operators $\mathcal{L}_{N-4}^*$ and $\mathcal{L}_{N-2}^*$
in the coordinates $X_1,Y_1;X_2,Y_2;\bx$ has the form
\begin{align}
\label{f-40}
\cL^{\star}_{N-4}v_{N-2}&=\cL^{\star}_{N-4}\bigg(\frac{Y_1-Y_2}{X_1-X_2}\bigg)=
\frac{2Y_2^2-2Y_1^2+(X_1-X_2)(P^{(1)}_{X_1}+P^{(2)}_{X_2})}{(X_1-X_2)^3},\\
\label{f-41}
\cL^{\star}_{N-4}v_{N}&=\cL^{\star}_{N-4}(Y_1+Y_2)=
\frac{P^{(1)}_{X_1}-P^{(2)}_{X_2}}{X_1-X_2}\,,\\
\cL^{\star}_{N-2}v_{N-2}&=\cL^{\star}_{N-2}\bigg(\frac{Y_1-Y_2}{X_1-X_2}\bigg)
=\frac{2(Y_2-Y_1)(X_1Y_1+X_2Y_1)+(X_1-X_2)(X_2P^{(1)}_{X_1}+X_1P^{(2)}_{X_2})}{
(X_1-X_2)^3}\,,\label{f-42}\\
\label{f-43}
\cL^{\star}_{N-2}v_{N}&=\cL^{\star}_{N-2}(Y_1+Y_2)=
\frac{X_2P^{(1)}_{X_1}-X_1P^{(2)}_{X_2}}{X_1-X_2}\,.
\end{align}
Our goal is to show that this action is polynomial in the coordinates
$u_2,u_4,v_{N-2},v_N,{\bf y}$.
We shall use the polynomials $a_{2k}(u_2,u_4)$
(see (\ref{f-32})) and the polynomials $b_{2n}(u_2, u_4)$ for which
$\sum b_{2n}\cdot t^n = a^2(t)$.

\begin{Lem}\label{L-17}
The action of the operators $\mathcal{L}_{N-4}^*$ and $\mathcal{L}_{N-2}^*$
on the coordinate functions
$v_{N-2}$ and $v_N$ has the form

\ \
\vglue-25pt
\begin{align}
\label{f-44}
\cL^{\star}_{N-4}v_{N-2}&=\sum_{k=0}^{N-3} (-1)^k y_{2k}b_{2N-2k-6},\\
\label{f-45}
\cL^{\star}_{N-4}v_{N}&= \sum_{k=0}^{N-1} (-1)^k (N-k)y_{2k}a_{2N-2k-4},\\
\cL^{\star}_{N-2}v_{N-2}&=v_{N-2}^2+\frac{1}{2}\,u_2 \sum_{k=0}^{N-3} (-1)^k 
y_{2k}
b_{2N-2k-6}-\frac{1}{2}\sum_{k=0}^{N-1} (-1)^k (N-k)y_{2k} 
a_{2N-2k-4},\label{f-46}\\
\label{f-47}
\cL^{\star}_{N-2}v_{N}&=\sum_{k=0}^{N-1} (-1)^k (N-k)y_{2k}(u_2a_{2N-2k-4} 
-a_{2N-2k-2}).
\end{align}
\end{Lem}

To prove this lemma, we need the following general statement.

\begin{Lem}\label{L-18}
The formula
\begin{equation}\label{r}
r(P)=\frac{2(P^{(2)}-P^{(1)})+(X_1-X_2)(P^{(1)}_{X_1}+P^{(2)}_{X_2})}{
(X_1-X_2)^3}
\end{equation}
defines a linear map
$r \colon \mathbb{C}[X;{\bf y}]\to\mathbb{C}[X_1;X_2;{\bf y}]$ of
$\mathbb{C}[{\bf y}]$-modules.
\end{Lem}

{\bf Poof} 
The transform (\ref{r}) is $\mathbb{C}[{\bf y}]$-linear; thus,
it suffices to prove that $r(X^k)\in \mathbb{C}[X_1,X_2;{\bf y}]$, 
$k=0,1,\dots$.
Let us take the generating series $f(t;X)=\sum_{k=0}^\infty X^kt^k=(1-tX)^{-1}$.
We obtain $r(f(t,X))=t^3 a^2(t)$, where $a(t)=a(t;u_2,u_4)$ (see
(\ref{f-32})).
Thus, we have $r(1)=r(X)=r(X^2)=0$ and $r(X^k)=b_{2k-6}$ for
$k\ge 3$, where the $b_{2n}$
are polynomials with generating series $\sum_{n=0}^\infty b_{2n}t^n = 
a^2(t)$.\\
\hfill $\Box $

We proceed to prove Lemma~\ref{L-17}.
Using (\ref{f-40}), we derive (\ref{f-44}).
Relation (\ref{f-46}) can be obtained by evaluating
$\cL^{\star}_{N-4}v_{N-2}$,
since (\ref{f-42}) can be rewritten as
$$
\cL^{\star}_{N-2}v_{N-2}=\bigg(\frac{Y_1-Y_2}{X_1-X_2}\bigg)^2+\frac{(X_1+X_2)}{
2}
\cL^{\star}_{N-4}v_{N-2}-\frac{1}{2}\bigg(\frac{P^{(1)}_{X_1}-P^{(2)}_{X_2}}{
X_1-X_2}\bigg),
$$
and applying the relation
$$
P_X= \sum_{k=0}^{N-1} (-1)^k (N-k)y_{2k}X^{N-k-1}.
$$
The expression (\ref{f-45}) for $\cL^{\star}_{N-4}v_{N}$
is obtained by using (\ref{f-34}).
Relation (\ref{f-43}) can be rewritten as
$$
\cL^{\star}_{N-2}v_{N}=\frac{1}{2}(X_1 + 
X_2)\cL^{\star}_{N-4}v_{N}-\frac{1}{2}(P^{(1)}_{X_1}+P^{(2)}_{X_2}).
$$
Again applying~(24),
we obtain (\ref{f-47}), which proves the lemma.\\

\smallskip

Thus, we have proved the following theorem, which is one of the main
results of the present paper.

\begin{The}\label{t-9}
For each $N\ge 3$, a Lie $\mathbb{C}[u_2,u_4,v_{N-2},v_N;{\bf y}]$-algebra with 
generators
${L}_{-2}^0,\dots,\break
{L}_{2(N-2)}^0,\mathcal{L}_{N-4}^*,\mathcal{L}_{N-2}^*$ is defined.
The commutation relations between these generators
are described in Corollaries~\ref{cor-12} and~\ref{cor-13}, and their action
on $u_2,u_4,v_{N-2},v_N,{\bf y}$, in Lemmas~\ref{L-13} and~\ref{L-17}.
\end{The}

\section{Examples of Polynomial Lie Algebras}

In this section
we give an explicit description of the polynomial Lie algebras $\mathcal{G}(N)$,
$N= 3, 4, 5$, over the rings
$\mathbb{C}[u_2,u_4,v_{1},v_3]$ for $N=3$,
$\mathbb{C}[u_2,u_4,v_2, v_4; y_4]$ for $N=4$, and
$\mathbb{C}[u_2,u_4,v_3, v_5; y_4, y_6]$ for $N=5$
with generators $\mathcal{L}_{0},\dots,\mathcal{L}_{2(N-2)},
\mathcal{L}_{N-4}^*,\mathcal{L}_{N-2}^*$. Here $L_0$ is the Euler field and, 
therefore,
$[L_0, L_{2k}]=2kL_{2k}$.

{\bf Proof} {Case $\boldsymbol{N=3}$}
We have
$$
y_{4}=\tfrac14(-3 u_{2}^2 - u_{4} + 4 v_{1} v_{3}),\quad
y_{6}=\tfrac14(-u_{2}^3 + u_{2} u_{4} - u_{4} v_{1}^2 + 2 u_{2} v_{1} v_{3}- 
v_{3}^2).
$$
The action of the generators $L_0$, $L_2$, and $\cL_{-1}^*$, $\cL_1^*$ of the 
free left $\bbbC[u_2,u_4,v_1,v_3]$-module is
as follows:
$$
\begin{tabular}{l|c|c|c|c|}
& $u_2$ & $u_4$ & $v_1$ & $v_3$ \\ \hline
$L_0$&$2u_2$&$4u_4$&$v_1$&$3v_3$\\
$L_2$&$\frac13 (3 u_2^2\,{-}\,u_4{-}\,8 v_1 v_3)$&$ -4 u_2 u_4
$&$ \frac12 (u_2 v_1{-}\,3v_3)$&$ -\frac32(u_4 v_1 + u_2 v_3)$\\
$\mathcal{L}_{-1}^*$&$2 v_1 $&$ 4 v_3 $&$ 1 $&$ 3 u_2$\\
$\mathcal{L}_1^*$&$u_2 v_1{-}\,v_3 $&$ -2 (u_4 v_1{-}\,u_2 v_3) $&$ -u_2 +
v_1^2 $&$ \frac12 (3 u_2^2\,{-}\,u_4{-}\,2 v_1 v_3)$
\end{tabular}
$$
The commutation relations are
$$
[\cL_{-1}^*,L_2]=-3 u_2\cL_{-1}^*+\cL_{1}^*,\qquad
[\cL_{1}^*,L_2]=\tfrac{1}{12} (9 u_4 -9 u_2^2 + 16
y_4)\cL_{-1}^*.
$$
\hfill $\Box $

{\bf Proof} {Case $\boldsymbol{N=4}$}
We have
\begin{gather*}
y_{6}=\tfrac14(2 u_{2}^3 + 2 u_{2} u_{4} - 4 v_{2} v_{4} + 4 u_{2} y_{4}),\\
y_{8}=\tfrac1{16}(3 u_{2}^4 - 2 u_{2}^2 u_{4} - u_{4}^2 + 4 u_{4} v_{2}^2 -
8 u_{2} v_{2} v_{4} + 4 v_{4}^2 + 4 u_{2}^2 y_{4} - 4 u_{4} y_{4}).
\end{gather*}
The action of the generators $L_0$, $L_2$, $L_4$, and $\cL^{\star}_{0}$, 
$\cL^{\star}_{2}$
of the free left $\bbbC[u_2,u_4,v_2,v_4,y_4]$-module is as follows:
$$
\begin{tabular}{l|c|c|c|}
& $y_4$&$u_2$&$u_4$ \\ \hline
$L_0$&$4y_4$&$2u_2$&$4u_4$\\
$L_2$& $6y_6$&$-u_{2}^2 - u_{4} - 2 y_{4}$ & $-4 u_{2} u_{4}$ \\
$L_4$&$8y_8$&$\frac12(u_{2}^3 + 3 u_{2} u_{4} + 4 u_{2} y_{4} - 6 y_{6})$ &
$u_{4} (3 u_{2}^2 + u_{4} + 4 y_{4})$\\
$\cL_{0}^*$&$0$&$2 v_2$ & $4 v_{4}$\\
$\cL_2^*$&$0$&$u_{2} v_2 - v_{4}$ & $-2 (u_{4} v_2 - u_{2} v_{4})$
\end{tabular}
$$
$$
\begin{tabular}{l|c|c|}
&$v_2$&$v_4$ \\ \hline
$L_0$&$2v_2$&$4v_4$\\
$L_2$ & $-2 v_{4}$ & $-2 (u_{4} v_2 + u_{2} v_{4})$ \\
$L_4$ & $\frac12 (-u_{2}^2 v_2 + u_{4} v_2 + 4 u_{2} v_{4})$ &
$2 u_{2} u_{4} v_2 + u_{2}^2 v_{4} + u_{4} v_{4} + 2 v_{4} y_{4}$\\
$\cL_{0}^*$& $2 u_{2}$ & $3 u_{2}^2 + u_{4} + 2 y_{4}$\\
$\cL_2^*$ & $\frac12 (-u_{2}^2 - u_{4} + 2 v_2^2 - 2 y_{4})$&$u_{2}^3 - u_{2} 
u_{4} + y_{6}$
\end{tabular}
$$
The commutation relations are
\begin{align*}
[L_{2},L_{4}]&=y_6L_0-y_4 L_2-(u_4 v_2+u_2 v_4)\cL^{\star}_{0}+2 v_4 
\cL^{\star}_{2},\\
[\cL^{\star}_{0},L_{2}]&=-2 u_2\cL^{\star}_{0},\\
[\cL^{\star}_{0},L_{4}]&=(3 u_2^2+u_4+2 y_4)\cL^{\star}_{0}-2 
u_2\cL^{\star}_{2},\\
[\cL^{\star}_{2},L_{2}]&=\tfrac12(u_4- u_2^2+2y_4)\cL^{\star}_{0},\\
[\cL^{\star}_{2},L_{4}]&=\tfrac12 (2 u_2^3 - 2 u_2 u_4 + 3 y_6)
\cL^{\star}_{0}-\tfrac12 (u_2^2-u_4) \cL^{\star}_{2}.
\end{align*}
\hfill $\Box $

{\bf Proof} {Case $\boldsymbol{N=5}$}
We have
\begin{align*}
y_{8}&=\tfrac1{16} (-5 u_{2}^4 - 10 u_{2}^2 u_{4} - u_{4}^2 + 16 v_{3} v_{5} -
12 u_{2}^2 y_{4} - 4 u_{4} y_{4} + 16 u_{2} y_{6}),\\
y_{10}&=\tfrac1{16}(-2 u_{2}^5 + 2 u_{2} u_{4}^2 - 4 u_{4} v_{3}^2 +
8 u_{2} v_{3} v_{5} \\
&\qquad\qquad- 4 v_{5}^2 - 4 u_{2}^3 y_{4}+
4 u_{2} u_{4} y_{4} + 4 u_{2}^2 y_{6} - 4 u_{4} y_{6}).
\end{align*}
The action of the generators $L_0$, $L_2$, $L_4$, $L_6$, and $\cL^{\star}_{1}$, 
$\cL^{\star}_{3}$
of the free left $\bbbC[u_2,u_4,v_3,v_5,y_4,y_6]$-module is as follows:
$$
\begin{tabular}{l|c|c|}
& $y_4$&$y_6$ \\ \hline
$L_0$&$4y_4$&$6y_6$\\
$L_2$& $6 y_{6}$& $-\tfrac45 (3 y_{4}^2 - 10 y_{8})$ \\[2pt]
$L_4$&$8 y_{8}$& $-\tfrac25 (4 y_{4} y_{ 6} - 25 y_{10})$\\[2pt]
$L_6$&$10 y_{10}$& $-\tfrac45 y_{4} y_{8}$\\[2pt]
$\cL_1^*$&$0$&$0$\\
$\cL_3^*$&$0$&$0$
\end{tabular}
$$
$$
\begin{tabular}{l|c|c|}
& $u_2$&$u_4$ \\ \hline
$L_0$&$2u_2$&$4u_4$\\
$L_2$& $\tfrac15 (-5 u_{2}^2 - 5 u_{4} - 8y_{4})$& $-4 u_{2} u_{4}$ \\[1pt]
$L_4$& $\tfrac1{10}(5 u_{2}^3 + 15u_{2} u_{4} + 20 u_{2} y_{4} - 24 y_{ 6})$&
$u_{4} (3 u_{2}^2 + u_{4} + 4 y_{4})$\\[2pt]
$L_6$&$\tfrac1{20}(-5 u_{2}^4 - 30 u_{2}^2 u_{4} - 5 u_{4}^2 - 20 u_{2}^2 
y_{4}$&$-2 u_{4}
(u_{2}^3 + u_{2} u_{4} + 2 u_{2}y_{4} - 2 y_{6})$\\
&$\kern40pt -20 u_{4} y_{4} + 40 u_{2} y_{6} - 64 y_{8})$&\\
$\cL_{1}^*$&$2 v_{3}$& $4 v_{5}$\\
$\cL_3^*$&$u_{2} v_{3} - v_{5}$& $-2 (u_{4} v_{3} - u_{2} v_{5})$
\end{tabular}
$$
$$
\begin{tabular}{l|c|}
&$v_3$ \\ \hline
$L_0$&$3v_3$\\
$L_2$ & $\tfrac12 (-u_{2} v_{3} - 5 v_{5})$\\[2pt]
$L_4$ & $\tfrac14 (10u_2v_5-(u_2^2-3v_4-4y_4)v_3)$\\[2pt]
$L_6$ & $\tfrac18 (3 u_{2}^3 v_{3} - 7 u_{2} u_{4} v_{3} - 15 u_{2}^2 v_{5} -
5 u_{4} v_{5} + 4 u_{2} v_{3} y_{4}-12 v_{5} y_{4})$ \\[2pt]
$\cL_1^*$& $\tfrac12 (5 u_{2}^2 + u_{4} + 2 y_{4})$\\
$\cL_3^*$ & $\tfrac12 (-2 u_{2} u_{4} + 2 v_{3}^2 - 2 u_{2} y_{4} + 2 y_{6})$
\end{tabular}
$$
$$
\begin{tabular}{l|c|}
&$v_5$ \\ \hline
$L_0$&$5v_5$\\
$L_2$ & $-\tfrac52(u_{4} v_{3} + u_{2} v_{5})$ \\[2pt]
$L_4$ & $\tfrac14 (10 u_{2} u_{4} v_{3} + 5 u_{2}^2 v_{5} +
5 u_{4} v_{5} + 12 v_{5} y_{4})$ \\[2pt]
$L_6$ & $\tfrac18 (-15 u_{2}^2 u_{4} v_{3}\,{-}\,5 u_{4}^2 v_{3}\,{-}\,
5 u_{2}^3 v_{5}\,{-}\,15 u_{2} u_{4} v_{5}\,{-}\,12 u_{4} v_{3} y_{4}\,{-}\,
12 u_{2} v_{5} y_{4}\,{+}\,16 v_{5} y_{6})$ \\[2pt]
$\cL_1^*$ & $\tfrac12 (5 u_{2}^3 + 5 u_{2} u_{4}+ 6 u_{2} y_{4} - 4 
y_{6})$\\[2pt]
$\cL_3^*$ & $\tfrac14 (5 u_{2}^4 - u_{4}^2 - 4 v_{3} v_{5} + 6 u_{2}^2 y_{4} - 
2 u_{4} y_{4} -
4 u_{2} y_{6})$
\end{tabular}
$$
The commutation relations are
\begin{align*}
[L_{2},L_{4}]&=2 L_6-\tfrac85 y_4L_2+\tfrac85 y_6L_0,\\
[L_{2},L_{6}]&=-4 v_5\cL^{\star}_{3}+2 (u_4 v_3+u_2v_5) 
\cL^{\star}_{1}-\tfrac45 y_4L_4+\tfrac45 y_8L_0,\\
[L_{4},L_{6}]&= (u_4 v_3+u_2 v_5)\cL^{\star}_{3}-\tfrac12 (2 u_2u_4 v_3+u_2^2 
v_5+u_4 v_5) \cL^{\star}_{1}+2 y_4L_6-\tfrac65 y_6L_4+\tfrac65y_8L_2-2 
y_{10}L_0,\\
[\cL^{\star}_{1},L_{2}]&=-\cL^{\star}_{3}- u_2\cL^{\star}_{1},\\
[\cL^{\star}_{1},L_{4}]&=- u_2\cL^{\star}_{3}+\tfrac14 (9 u_2^2+3u_4+4 
y_4)\cL^{\star}_{1},\\
[\cL^{\star}_{1},L_{6}]&=\tfrac14(9 u_2^2+3 u_4+4 y_4)\cL^{\star}_{3}-\tfrac12 
(5 u_2^3+5 u_2 u_4+6 u_2 y_4-4 y_6)\cL^{\star}_{1},\\
[\cL^{\star}_{3},L_{2}]&=\tfrac1{20}(-5 u_2^2+5 u_4+16 y_4)\cL^{\star}_{1},\\
[\cL^{\star}_{3},L_{4}]&=-\tfrac14(u_2^2-u_4+4 y_4)\cL^{\star}_{3}+\tfrac3{20} 
(5 u_2^3-5 u_2 u_4+8 y_6)\cL^{\star}_{1},\\
[\cL^{\star}_{3},L_{6}]&=-\tfrac1{80}(75 u_2^4 - 50 u_2^2 u_4 - 25 u_4^2 + 60 
u_2^2 y_4- 60 u_4 y_4 - 128 y_8) \cL^{\star}_{1}+\tfrac34 u_2 
(u_2^2-u_4)\cL^{\star}_{3}.
\end{align*}
\hfill $\Box $

\begin{The}\label{t-10}
There is an isomorphism of graded rings
$$
\varphi \colon \mathbb{C}[u_2,u_4,v_{3},v_5;y_4,y_{6}] \to 
\mathbb{C}[x_2,x_3,x_4;z_4,z_5,z_6],
$$
which determines an isomorphism of the polynomial Lie algebra described above 
(for $N=5$)
and the polynomial Lie algebra constructed in \cite{B-16} on the basis
of the theory of two-dimensional sigma-functions.
\end{The}

{\bf Poof} 
The isomorphism $\varphi$ and its inverse are given by
\begin{alignat*}2
u_{2}&\to x_2,&\qquad x_2&\to u_{2},\\[-2pt]
v_3&\to \tfrac12x_3,&\qquad x_3&\to 2 v_3,\\[-2pt]
u_{4}&\to x_2^2+4z_4,&\qquad x_4&\to 5 u_{2}^2+u_{4}+2 y_{4},\\[-2pt]
v_5&\to \tfrac12(x_2 x_3+2 z_5),&\qquad z_4&\to\tfrac{1}{4} 
(u_{4}-u_{2}^2),\\[-2pt]
y_{4}&\to \tfrac12(-6x_2^2+x_4-4 z_4),&\qquad z_5&\to v_5-u_{2} v_3,\\[-2pt]
y_{6}&\to -\tfrac14(-8 x_2 z_4+8 x_2^3-2 x_4x_2+x_3^2+2 z_6),&\qquad z_6&\to 2 
(u_{2}y_{4}+u_{2} u_{4}-v_3^2-y_{6}).
\end{alignat*}
A direct verification shows that this isomorphism determines the required 
isomorphism of polynomial
Lie algebras, which is the identity isomorohism
at the generators $L_0$, $L_2$, $L_4$, $L_6$, $\mathcal{L}_1^0$, and 
$\mathcal{L}_3^0$.
This proves the theorem.\\
\hfill $\Box $

\section{Integrable Hamiltonian polynomial   systems on $\bbbR^4$}\label{s10}

In the previous sections we have shown that a polynomial Lie algebra of vector 
fields can be canonically associated with a universal space of symmetric 
squares of  hyperelliptic curves. 
Commuting vector fields $\cL^{\star}_{N-4},\cL^{\star}_{N-2}$ correspond a 
pair of compatible dynamical systems in   variables 
$u_2,u_4,v_{N-2},v_N$ which depend on parameters $y_4,\ldots, y_{2N-4}$
\begin{equation}\label{dynsystems}
  \left\{\begin{array}{l}
       \partial_\xi u_2= \cL^{\star}_{N-4}u_2,\\
     \partial_\xi   u_4= \cL^{\star}_{N-4}u_4,\\
     \partial_\xi  v_{N-2}= \cL^{\star}_{N-4}v_{N-2},\\
      \partial_\xi v_N= \cL^{\star}_{N-4}v_{N};
        \end{array}\right.\qquad
 \left\{\begin{array}{l}
       \partial_\eta u_2= \cL^{\star}_{N-2}u_2,\\
     \partial_\eta   u_4= \cL^{\star}_{N-2}u_4,\\
     \partial_\eta  v_{N-2}= \cL^{\star}_{N-2}v_{N-2},\\
      \partial_\eta v_N= \cL^{\star}_{N-2}v_{N}.
        \end{array}\right. 
\end{equation}
The images $\cH_{2N-2}=\varphi(y_{2N-2})$  (\ref{phiy2N2}) and 
$\cH_{2N}=\varphi(y_{2N})$ (\ref{phiy2N})
are common first integrals of the 
above systems. Assuming the parameters $y_4,\ldots,y_{2N-4}$ and variables 
$u_2,u_4,v_{N-2},v_N$ to be real we obtain two integrable polynomial dynamical 
system on $\bbbR^4$. Moreover, the systems obtained are Hamiltonian with the 
Hamiltonians  $\cH_{2N-2},\cH_{2N}$ and the only nonzero Poisson brackets given 
by
\begin{equation}\label{poissonuv}
 \{u_2,v_N\}=2,\ \{u_4,v_{N-2}\}=4.
\end{equation}
These Hamiltonians are in involution since the corresponding vector fields 
commute. Thus the systems (\ref{dynsystems}) are Liouville integrable. In terms 
of the original coordinate $(X_1,Y_1;X_2,Y_2)\in \bbbC^4$  Poisson brackets 
(\ref{poissonuv}) correspond to the canonical brackets
\[
 \{X_i,Y_j\}=\delta_{ij}, \quad \{X_i,X_j\}=\{Y_i,Y_j\}=0.
\]

As we shall see below, in the cases  $N=3,4$ these systems can be 
integrated in terms of elliptic functions corresponding to the Jacobian of 
the curve. In the case $N=5,6$ the system can be integrated in Abelian 
functions.
For any $N\ne 5,6$ the system cannot be integrated in 
$2g=2\left[\frac{N-1}{2}\right]$ periodical 
Abelian functions. In the case $N=7$ a  general solution of the system can be 
expressed in terms of meromorphic functions which are 
$6$ periodic being restricted to the $\sigma$ 
divisor of the Jacobian \cite{bayano}.

\subsection{Dynamical system in the case $N=3$}

In the case $N=3$ the commuting vector fields $\cL^{\star}_{-1}, 
\cL^{\star}_{1}$ act on $\bbbR^4$ with coordinates $v_1,u_2,v_3,u_4$ and define 
two compatible (commuting) dynamical systems
\begin{eqnarray}
&& \left\{\begin{array}{l}\label{sys3_1}
 \partial_\xi v_1= 1,\\
     \partial_\xi   u_2 = 2 v_1,\\
     \partial_\xi  v_3 = 3 u_2,\\
      \partial_\xi u_4= 4 v_3;
        \end{array}\right.\\ \label{sys3_2}
&& \left\{\begin{array}{l}
       \partial_\eta {v}_1= v_1^2 - u_2 ,\\
        \partial_\eta {u}_2 = u_2 v_1 -v_3,\\
        \partial_\eta {v}_3=\frac{1}{2} (3 u_2^2 - u_4 - 2 v_1 v_3),\\
 \partial_\eta {u}_4 =  2 u_2 v_3-2u_4 v_1 ;
        \end{array}\right.
\end{eqnarray}
where $\partial_\xi$ and $\partial_\eta$ stand for $\cL^{\star}_{-1}$ and 
$\cL^{\star}_{1}$ respectively. The above dynamical systems possess two 
first integrals 
\begin{eqnarray}\label{y4N3}
 y_4 &=& v_1 v_3 - \frac{1}{4} (3 u_2^2 + u_4); \\ \label{y6N3}
 y_6 &=&  \frac{1}{4} (u_2 u_4-v_3^2 + 2 u_2 v_1 v_3 - u_2^3 - u_4 v_1^2).
\end{eqnarray}
in involution  (which are the Hamiltonians of these systems) with respect to 
the Poisson structure given by (\ref{poissonuv})
and thus are Liouville integrable. Moreover it is not difficult to 
show 
that a general simultaneous solution to the systems can be written in the form
\begin{equation}\label{uvN3}
  \begin{array}{l}
  v_1=\xi+f,\\
  u_2=\xi^2+2\xi f+f^2-f_\eta ,\\
  v_3=\xi^3+3\xi^2 f+3\xi (f^2-f_\eta)+f^3-3f_\eta f+f_{\eta\eta},\\
   u_4=\xi^4+4\xi^3 f+6\xi^2 (f^2-f_\eta)+4\xi (f^3-3f_\eta 
f+f_{\eta\eta})+g.\\
 \end{array}
\end{equation}
where 
\[
 g=f^4-6f_\eta 
f^2+9f_\eta^2+4f_{\eta\eta}f-2f_{\eta\eta\eta}.
\]
and function $f=f(\eta)$ is a general solution to the equation
\[
f_{\eta\eta} ^2  = f_\eta^3 +4 y_4 f_\eta -4y_6
\]
as it follows from system 
(\ref{sys3_2}) and the first integral (\ref{y6N3}). A general solution of this 
equation can be expressed in terms of the Weierstrass $\zeta$ function
\begin{equation}\label{fN3}
 f=-\zeta(\eta-\eta_0;g_2,g_3)+f_0,\qquad g_2=-4y_4,\quad  
g_3=4y_6.
\end{equation}
Thus (\ref{uvN3}) together with (\ref{fN3}) represent a simultaneous general 
solution to the system (\ref{sys3_1}), (\ref{sys3_2}) which depends on four 
arbitrary parameters $f_0,\eta_0,y_4,y_6$.

\subsection{Dynamical system in the case $N=4$}

In the case $N=4$ we consider dynamical systems corresponding the symmetric 
square of elliptic curve
\begin{equation}\label{curve4}
 Y^2=X^4+y_4 X^2-y_6 X+y_8.
\end{equation}
The commuting vector fields $\cL^{\star}_{0}, 
\cL^{\star}_{2}$ act on $\bbbR^4$ with coordinates $u_2,v_2,u_4,v_4$ and define 
two compatible (commuting) dynamical systems depending on the parameter $y_4$
\begin{equation}\label{sys4}
 \left\{\begin{array}{l}
       \partial_\xi u_2= 2v_2,\\
     \partial_\xi   v_2 = 2u_2 ,\\
     \partial_\xi  u_4 = 4v_4,\\
      \partial_\xi v_4=3 u_2^2 + u_4 + 2 y_4 ;
        \end{array}\right.\qquad
 \left\{\begin{array}{l}
       \partial_\eta {u}_2=  u_{2} v_{2} - 
 v_{4} ,\\
        \partial_\eta {v}_2 = \frac{1}{2} (-u_{2}^2 - u_{4} + 
   2 v_{2}^2 - 2 y_{4})  ,\\
        \partial_\eta {u}_4=2  u_{2} v_{4}-2u_{4} v_{2}  ,\\
 \partial_\eta {v}_4 = \frac{1}{2} (3 u_{2}^3 - u_{2} u_{4} - 2 v_{2} v_{4} + 2 
u_{2} y_{4})  ;
        \end{array}\right.
\end{equation}
where $\partial_\xi$ and $\partial_\eta$ stand for $\cL^{\star}_{0}$ and 
$\cL^{\star}_{2}$ respectively. The above dynamical systems possess two common 
first integrals (Hamiltonians) 
\begin{equation}\label{fint68}
 \begin{array}{ll}
y_{6}&= \frac{1}{4}(2 u_{2}^3 + 2 u_{2} u_{4} - 4 v_{2} v_{4}  + 4 u_{2} 
y_{4}),\\ &\\
y_{8}&= \frac{1}{16}(3 u_{2}^4 - 2 u_{2}^2 u_{4} - u_{4}^2 + 4 u_{4} v_{2}^2 - 
    8 u_{2} v_{2} v_{4} + 4 v_{4}^2   + 4 u_{2}^2 y_{4} - 4 u_{4} y_{4}) 
\end{array}
\end{equation} 
in invoution 
and thus are Liouville  integrable.

The first system is linear inhomogeneous with constant coefficients and its 
general solution is of the form
\begin{equation}\label{sys1sol4}
 \begin{array}{l}
  u_2=\alpha_1 e^{2\xi}+\alpha_2 e^{-2\xi},\\
  v_2=\alpha_1 e^{2\xi}-\alpha_2 e^{-2\xi},\\
  u_4=\alpha_1^2 e^{4\xi}+\alpha_2^2 
e^{-4\xi}-6\alpha_1\alpha_2-2y_4+2\beta_1 e^{2\xi}+2\beta_2 e^{-2\xi},\\
v_4=\alpha_1^2 e^{4\xi}-\alpha_2^2 
e^{-4\xi}+ \beta_1 e^{2\xi}-\beta_2 e^{-2\xi},
 \end{array}
\end{equation}
where $\alpha_1,\alpha_2,\beta_1,\beta_2$ are arbitrary constants. In order to 
satisfy the second system we should assume that 
$\alpha_1,\alpha_2,\beta_1,\beta_2$ are functions of 
$\eta$. It follows from the second system that 
\begin{eqnarray}\label{bi}
&&\alpha_1' = -\beta_1,\qquad\qquad\qquad\    \alpha_2' = \beta_2, \\
&&\beta_1' = 2 \alpha_1  (4 \alpha_1  \alpha_2 +y_4),\quad \beta_2' = -2 
\alpha_2 (4 \alpha_1  \alpha_2 +y_4).
\end{eqnarray}
Variables $\beta_1,\beta_2$ can be eliminated
\begin{equation}\label{sysalpha}
 \begin{array}{l}
  \alpha_1'' + 2 \alpha_1  (4 \alpha_1  \alpha_2 +y_4)=0,\\ 
  \alpha_2'' +2\alpha_2 (4 \alpha_1  \alpha_2 +y_4)=0.
 \end{array}
\end{equation}
Substitution (\ref{sys1sol4}), (\ref{bi}) in  (\ref{fint68}) results in two  
first integrals
\begin{equation}\label{fialpha}
 \begin{array}{l}
 y_8= \alpha_1'  \alpha_2' +\frac{1}{4} (4 \alpha_1  
\alpha_2 +y_4)^2,\\y_6= 2 \left(\alpha_1  \alpha_2' -\alpha_2  
\alpha_1' \right)
 \end{array}
\end{equation}
of the system. They can be viewed as a result of two integrations of the system 
(\ref{sysalpha}) with two arbitrary constants $y_6,y_8$. Equations 
(\ref{fialpha}) is a system of two first order equations which can be 
integrated using elliptic functions. Indeed, let us introduce function 
$r=\alpha_1\alpha_2$. Then, it follows from equations (\ref{fialpha}) that 
\begin{equation}\label{req4}
 r_\eta^2+16r^3+8y_4 r^2+(y_4^2-4y_8)r-\frac{1}{4}y_6^2=0.
\end{equation}
We can transform this equation to the canonical Weierstrass form
 by a linear change of variables
\[\rho=-4 r-\frac{2}{3}y_4,\]
then  (\ref{req4}) takes the form
\begin{equation}\label{rho4}
 (\rho')^2=4\rho^3-g_2 \rho -g_3,\qquad g_2=16y_8+\frac{4}{3}y_4^2,\quad 
g_3=\frac{32}{3}y_4y_8-4y_6^2-\frac{8}{27}y_4^3.
\end{equation}
It is well known (see for example \cite{WhittWat}) that the modular invariants 
$\hat g_2,\hat g_3$ of a regular elliptic  curve
\[
 Y^2=a_0X^4 + 4a_1X^3 + 6a_2X^2 +4a_3X +a_4
\]
are of the form
\[
 \hat g_2=a_0a_4 -4a_1a_3 +3(a_2)^2,\quad
\hat g_3 = a_0a_2a_4 +2a_1a_2a_3   -(a_2)^3 - a_0(a_3)^2 - (a_1)^2a_4.
\]
In our case $a_0=1,\ a_1=0,\ a_2=y_4/6,\ a_3=-y_6/2,\ a4 =y_8$ and thus 
\[   g_2=4^2 \hat g_2,\quad   g_3=4^3 \hat g_3\]
and therefore the modules  of the original curve (\ref{curve4}) 
and the curve corresponding to equation (\ref{rho4}) coincide.
Solution to the above equation can be expressed in terms of the 
Weierstrass elliptic function 
\[
 \rho = \wp (\eta-\eta_0;g_2,g_3),
\]
and therefore
\[
r=-\frac{1}{4}\wp (\eta-\eta_0;g_2,g_3)-\frac{1}{6}y_4
\]
where $\eta_0$ is a constant of integration. The second equation in the system 
(\ref{fialpha}) can be rewritten in the form
\[
 2\left(\ln \frac{\alpha_2}{\alpha_1}\right)_\eta =\frac{y_6}{r}
\]
and integrated
\[
 \frac{\alpha_2}{\alpha_1}=C \exp\left( 
y_6\int\frac{d\eta}{2r}\right)
\]
where $C$ is an arbitrary integration constant. Thus (\ref{sys1sol4}) 
with 
\[
\alpha_1=\sqrt{C^{-1}r} 
\exp\left( -y_6
\int\frac{d\eta}{4r}\right),\quad 
\alpha_2= 
\sqrt{C r } 
\exp\left(   y_6 
\int\frac{d\eta}{4r}\right)
\]
and $\beta_1=-\alpha_{1\eta},\beta_2=\alpha_{2\eta}$ is a general solution of 
 two compatible systems  (\ref{sys4}) with four arbitrary constants
 $\eta_0,C,y_6,y_8$ and an arbitrary parameter $y_4$.

\subsection{Dynamical system in the case $N=5$}

In the case $N=5$ the commuting vector fields $\cL^{\star}_{1}, 
\cL^{\star}_{3}$ (which represent $\partial_\xi$ and $\partial_\eta$)  act on 
$\bbbR^4$ with coordinates $u_2,v_3,u_4,v_5$ and define 
two compatible (commuting) dynamical systems
\begin{eqnarray}\label{sys5_1}
&&
\left\{\begin{array}{l}
       \partial_\xi u_2= 4 u_{3},\\
     \partial_\xi   v_3 =  5 u_{2}^2 + u_{4} + 2y_{4},\\
     \partial_\xi  u_4 = 8u_{5},\\
      \partial_\xi v_5=5 u_{2}^3 + 5 u_{2} u_{4} + 6 u_{2} y_{4} - 4 y_{6};
        \end{array}\right.
\\ \nonumber
\\ \label{sys5_2}
&& \left\{\begin{array}{l}
\partial_\eta {u}_2= 4  u_{5}-4u_{2} u_{3}  ,\\
\partial_\eta {v}_3 = -4 (u_{3}^2 - u_{2}u_{4} -u_{2} y_{4} + y_{6}),\\
        \partial_\eta {u}_4= 8 (u_{3} u_{4} - u_{2}u_{5}),\\
 \partial_\eta {v}_5 =  u_{4}^2 -5 u_{2}^4 +
4 u_{3} u_{5} - 6 u_{2}^2 y_{4} + 2 u_{4} y_{4} + 4 u_{2}y_{6};
        \end{array}\right.
\end{eqnarray}
with two common first integrals
\begin{eqnarray*}
&&y_{8}=\tfrac1{16} (-5 u_{2}^4 - 10 u_{2}^2 u_{4} - u_{4}^2 + 16 v_{3} v_{5} -
12 u_{2}^2 y_{4} - 4 u_{4} y_{4} + 16 u_{2} y_{6}),\\
&&y_{10}=\tfrac1{16}(-2 u_{2}^5 + 2 u_{2} u_{4}^2 - 4 u_{4} v_{3}^2 +
8 u_{2} v_{3} v_{5} - 4 v_{5}^2 - 4 u_{2}^3 y_{4}+
4 u_{2} u_{4} y_{4} + 4 u_{2}^2 y_{6} - 4 u_{4} y_{6}).
\end{eqnarray*}

This system can be integrated in terms of Abelian functions on the Jacobian. 
Indeed, consider the sigma-function $\sigma=\sigma(w;\widehat y)$, 
$w=(\xi,\eta)$ (see 
\cite{BEL-97-1})
associated with the curve
\begin{equation}\label{f-53}
\{ (X,Y)\in\mathbb{C}^2: Y^2=X^5+y_4X^3-y_6X^2+y_8X-y_{10} \},
\end{equation}
and define Abelian $\wp_{i,3j}$ functions \cite{B-16}
$$
\wp_{i,3j}=\wp_{i,3j}(w;\widehat y)=-\frac{\partial^{i+j}}{\partial 
\xi^i\partial \eta^j}\ln\sigma.
$$
Then a general common solution for the two compatible systems can be written in 
the form:
\[
 \begin{array}{ll}
u_2=\wp_{2,0}(w-w_0;\widehat y),& v_3=\frac{1}{2}\wp_{3,0}(w-w_0;\widehat y),\\ 
& \\
u_4=\wp_{2,0}^2(w-w_0;\widehat y)+4\wp_{1,3}(w-w_0;\widehat y),\ & 
v_5=\frac{1}{2}\wp_{2,0}(w-w_0;\widehat y)\wp_{3,0}(w-w_0;\widehat 
y)+\wp_{2,3}(w-w_0;\widehat y).
 \end{array}
\]
The above solution depends on four arbitrary constants 
$w_0=(\xi_0,\eta_0),y_8,y_{10}$. Function $u_2$ satisfies the classical KdV 
equation and represents two gap solution \cite{B-16}. In  variables 
$X_1,Y_1,X_2,Y_2$  (Lemma \ref{L-9}) systems (\ref{sys5_1}) and  (\ref{sys5_2}) 
 correspond to Dubrovin's systems   (2.12) and  (3.9) in  \cite{dubrovin75}
respectively. Thus our approach can be considered
as a generalisation of the Dubrovin construction to the cases when the 
symmetric power of the curve does not coincide with its genus. Moreover, we 
have shown that in the case g = 2 a change of variables defined in Lemma 
\ref{L-9} transforms Dubrovin's systems 
in polynomial Hamiltonian systems.  It is important that in our 
approach the problem to construct real valued solutions can be easily resolved, 
indeed it is sufficient to choose the parameters of the curve $y_4,\ldots 
,y_{10}$ to be real.

\subsection{Dynamical system in the case $N=7$}

In the case $N=7$ the commuting vector fields $\cL^{\star}_{3}, 
\cL^{\star}_{5}$ (which represent $\partial_\xi$ and $\partial_\eta$)  act on 
$\bbbR^4$ with coordinates $u_2,u_4,v_5,v_7$ and define 
two compatible (commuting) dynamical systems

\begin{equation*}
\begin{array}{l}
\left\{\begin{array}{lll}
       \partial_\xi u_2&=& 2v_5 ,\\
     \partial_\xi   u_4 &=&4 v_7  ,\\
     \partial_\xi  v_5 
&=&\frac{1}{16}(40 u_{2}^2 y_{4}-32 u_{2} 
y_{6}+8 u_{4} y_{4}+35 u_{2}^4+42 u_{4} u_{2}^2+3 u_{4}^2+16 y_{8} ),\\
      \partial_\xi v_7&=&\frac{1}{16}(40 u_{2}^3 y_{4}-48 u_{2}^2 y_{6}+40 
u_{4} u_{2} y_{4}+48 u_{2} y_{8}-16 u_{4} 
y_{6}+21 u_{2}^5+70 u_{4} u_{2}^3+21 u_{4}^2 u_{2}-32 y_{10}) ;
        \end{array}\right.
\\
\\
 \left\{\begin{array}{lll}
\partial_\eta {u}_2&=& u_{2} v_{5}-v_{7}   ,\\
\partial_\eta {u}_4 &=&-2 u_{4} v_{5}+2u_{2}v_{7} ,\\
        \partial_\eta {v}_5&=& 
\frac{1}{16}   (   8 u_{2}^2 y_{6}-16 u_{4} u_{2} y_{4}-16 u_{2} y_{8}+8 
u_{4} y_{6}+7 u_{2}^5-14 u_{4} u_{2}^3-9 u_{4}^2 u_{2}+16 v_{5}^2+16 
y_{10}),\\
 \partial_\eta {v}_7 &=&  
\frac{1}{32}  (   40 u_{2}^4 y_{4}-48 u_{2}^3 y_{6}+48 u_{2}^2 y_{8}+16 
u_{4} u_{2} y_{6}-32 u_{2} y_{10}-8 u_{4}^2 y_{4}-16 u_{4} y_{8}+\\
&&21 
u_{2}^6+35 u_{4} 
u_{2}^4-21 u_{4}^2 u_{2}^2-3 u_{4}^3-32 v_{5} v_{7} );
        \end{array}\right.
\end{array}
\end{equation*}
These two compatible systems possess two common first integrals:
\begin{eqnarray*}
y_{12}&=& \frac{1}{64}
  (-7 u_{2}^6 - 35 u_{2}^4 u_{4} - 21 u_{2}^2 u_{4}^2 - u_{4}^3 + 
    64 v_{5} v_{7} + 12 u_{2}^5 y_{2} + 40 u_{2}^3 u_{4} y_{2} + 
    12 u_{2} u_{4}^2 y_{2} -\\&&- 20 u_{2}^4 y_{4} - 40 u_{2}^2 u_{4} y_{4} - 
    4 u_{4}^2 y_{4} + 32 u_{2}^3 y_{6} + 32 u_{2} u_{4} y_{6} - 
    48 u_{2}^2 y_{8} - 16 u_{4} y_{8} + 64 u_{2} y_{10}),\\
\\
y_{14}&=&\frac{1}{64}(-3 u_{2}^7 - 7 u_{2}^5 u_{4} + 7 u_{2}^3 u_{4}^2 + 3 
u_{2} u_{4}^3 - 
    16 u_{4} v_{5}^2 + 32 u_{2} v_{5} v_{7} - 16 v_{7}^2 + 
    5 u_{2}^6 y_{2} +\\&&+ 5 u_{2}^4 u_{4} y_{2} - 9 u_{2}^2 u_{4}^2 y_{2} - 
    u_{4}^3 y_{2} - 8 u_{2}^5 y_{4} + 8 u_{2} u_{4}^2 y_{4} + 
    12 u_{2}^4 y_{6} - 8 u_{2}^2 u_{4} y_{6} - 4 u_{4}^2 y_{6} - 
    16 u_{2}^3 y_{8} + \\&&+16 u_{2} u_{4} y_{8} + 16 u_{2}^2 y_{10} - 
    16 u_{4} y_{10}).
\end{eqnarray*}

For the hyperelliptic curve of genus 3 
\begin{equation}\label{cg3}
 V_{\bf y} = \{ (X,Y) \in \mathbb{C}^2\colon Y^2 = X^7 + y_4 X^5 + \ldots 
-y_{14}\},
\end{equation}
there defined $\sigma$ function $\sigma(w_1,w_3,w_5)$ (see \cite{BEL-97-1}). It 
is an 
entire function on $\bbbC^3$ with co-ordinates $w_1,w_3,w_5$. In $\bbbC^3$ 
there is an analytic surface $W=\{(w_1,w_3,w_5)\in\bbbC^3\,|\, 
\sigma(w_1,w_3,w_5)=0\}$. The surface $W$ is 6-periodic in $\bbbC^3$ and thus 
provides us with a $\sigma$--divisor, namely 
\[
 W\diagup \Lambda\subset {\rm Jac}(V),
\]
where $\Lambda$ is the lattice of periods.

The above system can be integrated in  meromorphic functions on $\bbbC^3$
which can be explicitly expressed 
 in terms of the gradient of the sigma function \cite{bayano}. The 
solutions of the above system  are meromorphic on $\bbbC^3$, they are not 
Abelian with respect to the lattice $\Lambda$ of the curve (\ref{cg3}), but 
their 
restrictions on $W$  are 6-periodic and therefore correctly defined on 
the $\sigma$ divisor.

\section*{Acknowledgements}
We are grateful to  V.~M.~Rubtsov, V.~V.~Sokolov and
A.~V.~Tsiganov,    for useful discussions of the results of our work.

\bibliography{kdv}
\end{document}